\documentclass[12pt,a4paper]{article}
\pdfoutput=1
\usepackage{graphicx,epsfig}
\usepackage{dcolumn} 
\usepackage{slashed,color,amsmath,amssymb}
\usepackage{jheppub}
\usepackage{enumitem}
\usepackage{siunitx}
\usepackage{bm}
\usepackage{layouts}
\usepackage[svgnames]{xcolor}
\usepackage{float}
\usepackage{caption, subcaption}
\usepackage{multirow}
\usepackage{verbatim}
\usepackage[normalem]{ulem}
\newcommand{\be}{\begin{equation}}
\newcommand{\ee}{\end{equation}}
\newcommand{\bea}{\begin{eqnarray}}
\newcommand{\eea}{\end{eqnarray}}
\newcommand{\bit}{\begin{itemize}}
\newcommand{\eit}{\end{itemize}}

\newcommand{\lam}{\lambda}

\definecolor{greeen}{HTML}{008ae6}

\definecolor{newgreen}{HTML}{009900}
\definecolor{newnewgreen}{HTML}{21db2c}

\newcommand\thefontsize{The current font size is: \f@size pt}

\newcommand{\no}[1]{{\color{red}{#1}}}
\newcommand{\yes}[1]{{\color{newgreen}{#1}}}

  \definecolor{fsred}{HTML}{c71616}

\definecolor{fsblue}{HTML}{1e88e5}

\definecolor{fsyellow}{HTML}{dca607}

\definecolor{fsgreen}{HTML}{014d40}

\definecolor{fsviolet}{HTML}{a11b9f}

\definecolor{newpurple}{HTML}{7B1DAA}

\def\gsim{\lower0.5ex\hbox{$\:\buildrel >\over\sim\:$}}
\def\lsim{\lower0.5ex\hbox{$\:\buildrel <\over\sim\:$}}

\hypersetup{
   colorlinks=true,       
   linkcolor=blue,        
   citecolor=red,         
   filecolor=magenta,     
}

\preprint{\begin{flushright} BONN-TH-2025-08
	\end{flushright}}	

\graphicspath{{./figs/}}

\title{The ABC of RPV II: Classification of R-parity Violating 
Signatures from UDD Couplings and their Coverage at the LHC}

    \author[a]{Herbi\,K.\,Dreiner,}
	\emailAdd{dreiner@uni-bonn.de}
	\affiliation[a]{Bethe Center for Theoretical Physics \& Physikalisches Institut der Universit\"at Bonn,\\ Nu{\ss}allee 12, 53115 Bonn, Germany}

    \author[b]{Michael Hank,}
    \emailAdd{mhank@sas.upenn.edu}
    \affiliation[b]{Department of Physics, University of Pennsylvania, Philadelphia PA; USA}
  
	\author[c]{Yong Sheng Koay,}		 
    \emailAdd{yongsheng.koay@physics.uu.se}
    \affiliation[c]{Department of Physics and Astronomy, Uppsala University, Sweden}		
		
    \author[a]{Martin Sch\"urmann,}
	\emailAdd{marschu@uni-bonn.de}
					
	\author[a]{Rhitaja Sengupta,}
	\emailAdd{rsengupt@uni-bonn.de}
		
	\author[a]{Apoorva Shah,}
	\emailAdd{ashah@uni-bonn.de}

    \author[d]{Nadja Strobbe,}
	\emailAdd{nstrobbe@umn.edu}
    \affiliation[d]{School of Physics \& Astronomy, University of Minnesota, Minneapolis, MN 55455, USA}
    
    \author[b]{and Evelyn Thomson}
    \emailAdd{thomsone@physics.upenn.edu}

\abstract{
We perform a detailed study of the current 
phenomenological status of baryon number violating 
operators within the framework of the $R$-parity violating 
Minimal Supersymmetric Standard Model (RPV-MSSM). This study 
aims to identify any gaps in the experimental coverage of 
the RPV landscape. We identify the unique final states for all 
possible LSPs decaying via four different benchmark UDD 
operators. Both the direct production of the LSP and its 
production via gauge-cascades are considered. For each LSP, we 
assume that only one UDD coupling is non-zero at a time and 
confront the signals with existing ATLAS and CMS searches 
implemented in the recasting framework \texttt{CheckMATE\;2}.
We find that the UDD colored LSP sector is well covered 
with the mass bounds on the gluino LSP being the strongest, and with
possible improvements for some of the right-handed squark LSPs. 
We also point out that there is limited coverage for 
electroweakino and slepton LSPs with UDD decays. This limitation
may be due to the lack of targeted experimental searches for 
these specific final states or the appropriate recasting of 
existing searches.
}

\begin{document}
\maketitle



\section{Introduction}
\label{sec:intro}

The supersymmetric (SUSY)\,\cite{Golfand:1971iw, 
Volkov:1973ix, Wess:1973kz, Wess:1974tw} extension of the
Standard Model of particle physics (SM) has been in the limelight for a long time. It is a
well-motivated beyond-the-SM (BSM) theory, which can 
offer solutions to several drawbacks of having 
the SM as the sole theory, such as the hierarchy problem\,
\cite{Weinberg:1975gm, Gildener:1976ih, Gildener:1976ai, 
Susskind:1978ms, Veltman:1980mj, tHooft:1980xss}. The most widely 
discussed supersymmetric model is the ``Minimal Supersymmetric Standard Model'' (MSSM) 
\cite{Drees:2004jm, Djouadi:2005gj, Baer:2006rs, 
Aitchison:2007fn, Dreiner:2023yus}, where the SM particle 
content is extended by one Higgs 
doublet and then supersymmetrized. Often, the 
conservation of an extra $\mathbb{Z}_2$ symmetry, called 
$R$-parity\,\cite{Farrar:1978xj,Barbier:2004ez}, is 
imposed. The conservation of $R$-parity eliminates 
baryon- and lepton-number violation 
from the renormalizable superpotential. 
Conservation of $R$-parity also means that supersymmetric 
particles can only be produced or annihilated in pairs. 
Furthermore, the 
lightest supersymmetric particle (LSP) is then 
necessarily stable and is an attractive candidate for dark matter  
\cite{Goldberg:1983nd}. In this case, it must be 
neutral with respect to electric charge and color. The 
collider signatures of the $R$-parity conserving MSSM 
(RPC-MSSM), therefore, involve final states with large
missing transverse momentum (which is also known as the missing transverse energy, $E_T^{\rm miss}$), due to 
LSP being invisible to the detector. Being a popular BSM theory, this 
scenario has been tested extensively in experiments, 
especially at the Large Hadron Collider (LHC) by the 
ATLAS and CMS collaborations. The 
extensive searches at colliders put severe 
constraints on the RPC-MSSM, pushing the lower bounds on 
the mass of SUSY particles in the colored sector to 
around 1-2\,TeV, and in the 
electroweak sector to around a few hundreds of GeV to a 
TeV\,\cite{ATLAS:2022ihe, ATLAS:2022ckd, ATLAS:2021jyv, 
ATLAS:2021hza,ATLAS:2021yij, ATLAS:2021twp,ATLAS:2020xzu,
ATLAS:2020syg,ATLAS:2020xgt,ATLAS:2020dsf,ATLAS:2019fag, 
ATLAS:2019gdh, CMS:2022vpy, CMS:2021eha, CMS:2021edw, 
CMS:2020bfa, CMS:2020cpy,CMS:2020fia, CMS:2020cur, 
CMS:2019lrh,CMS:2019ybf,CMS:2019zmd, CMS:2022idi,
ATLAS:2022hbt,ATLAS:2021yqv, ATLAS:2021moa, 
ATLAS:2019wgx, ATLAS:2019lng, ATLAS:2019gti, 
ATLAS:2020pgy,CMS:2021cox, CMS:2017moi, CMS:2021few,
CMS:2018eqb,CMS:2022syk,CMS:2019zmn, CMS:2019pov, 
CMS:2022sfi, CMS:2019eln}, depending on the 
model assumptions.

By omitting the imposed $\mathbb{Z}_2$ symmetry, the most general MSSM superpotential includes terms which violate $R$-parity, 
known as the $R$-parity violating MSSM (RPV-MSSM) 
\cite{Hall:1983id, Allanach:2003eb}. The RPV terms are 
usually set to zero by the imposition of RPC since these 
couplings can lead to proton decay at rates excluded by 
experiment\,\cite{Allanach:1999ic, Dreiner:2003hw, 
Chemtob:2004xr,Barbier:2004ez,Chamoun:2020aft, 
ParticleDataGroup:2024cfk}. However, this is not true 
when only some of the RPV couplings are present 
\cite{Ibanez:1991pr,Dreiner:2003yr, Dreiner:2005rd, 
Dreiner:2006xw, Dreiner:2012ae} and if the values of 
these couplings are below the experimental bounds\,
\cite{Barbier:2004ez,ParticleDataGroup:2024cfk,Domingo:2024qoj}. 
Therefore, there is no compelling reasoning to set all 
these terms to zero\,\cite{Dreiner:1997uz}. 

Moreover, the RPV-MSSM leads to a richer phenomenology 
than the RPC-MSSM, where the latter is mostly 
characterized by large $E_T^{\rm miss}$. The LSP in 
RPV-MSSM is not stable and can decay to SM particles, 
depending on which of the RPV couplings is non-zero. 
It is thus not a viable dark matter candidate and
also not  restricted to  being electrically or color 
neutral \cite{Dreiner:2008ca, Dercks:2017lfq}. The final
state, therefore, does not always consist of $E_T^{\rm 
miss}$. Signatures of the LSP decay in RPV-MSSM 
have been explored in a wide variety of 
phenomenological studies, see Refs.\,\cite{Dreiner:1994tj,Dreiner:1996dd,
RparityWorkingGroup:1999ixj,Datta:2000yc,Desai:2010sq,Bhattacherjee:2011dt,
Bhattacherjee:2013gr,Dercks:2017lfq,Domingo:2018qfg,
Bansal:2018dge,Dreiner:2020lbz,Barman:2020azo,
FileviezPerez:2022ypk,Dib:2022ppx,Choudhury:2023eje,
Choudhury:2023yfg,Choudhury:2024ggy,Domingo:2024qoj,
Choudhury:2024crp,Bickendorf:2024ovi}.
Ref.\,\cite{Dreiner_2023} demonstrates the wide variety 
of collider signatures arising in RPV-MSSM, see also
Refs.\,\cite{Dreiner:2012wm, Dercks:2017lfq}. It provides a
dictionary for the RPV-MSSM signatures depending on the 
various LSPs and the different RPV couplings. Such a 
systematic representation of the final states is useful 
to identify the relevant experimental searches. This, in 
turn, reveals any gaps in the present experimental 
searches.

In Ref.\,\cite{Dreiner_2023}, the status of the LLE operators ($\frac{1}{2} \lam_{ijk} L_i L_j \bar{E}_k$ operators in the MSSM superpotential) after LHC Run-2 was studied in detail for the different LSPs in RPV-MSSM.
This work aimed to answer the 
question: are the above bounds robust, or are there 
gaps/loopholes that could still allow LHC-scale SUSY to
be hiding? In the present work, we extend the previous 
analysis and perform a similarly detailed study for the UDD 
operators: $\frac{1}{2} \lam''_{ijk} \bar{U}_i \bar{D}_j 
\bar{D}_k$, in the RPV-MSSM superpotential. While 
classifying the signatures of the various LSP decays for 
the UDD couplings, $\lam''$, the previous study identified 
that the experimental coverage for the UDD operators might 
be less comprehensive than in the LLE scenario. Therefore, 
we perform a detailed numerical study of the UDD benchmark 
operators, which cover the full set of the possible final 
states for a range of LSPs, either produced directly or 
from the cascade decay of another supersymmetric particle 
at the LHC. A similar study can be performed to cover the final states of the LQD operators. Due to the large number of LQD operators (27), we leave this for an upcoming study.

The motivation for this work is to identify whether there 
are gaps in the experimental coverage or loopholes in the 
analyses performed for the baryon number violating couplings.
With the ongoing LHC Run-3, it is timely to 
assess any gaps or loopholes, and to improve the 
sensitivity for the UDD operators in the RPV-MSSM. The 
reason for the
existence of gaps in the coverage might be two-fold. First, there might be an 
experimental analysis that was originally aimed at a 
different BSM model but has similar final states as the signature of some LSP decaying through a UDD operator. The result of this search then has to be properly recast for the UDD operators to study if an improved sensitivity can be achieved. Even if the 
experimental result is available for the final state under study arising from a similar model,
it is mostly presented with specific assumptions, and a recasting is required to study the sensitivity for a different set of assumptions. In numerous studies, the information made publicly available by the experimental collaborations is either not adequate or not present in a form that can be directly used for recasting. Identifying such analyses and highlighting the useful information required for recasting is thus beneficial for both experimentalists and theorists. 

The second possibility for the existence of a gap in the coverage might be the lack of a sensitive analysis for some of the final states at the LHC. In this case, they can be a focal point of Run-3. With the present work, we want to point out gaps where a proper recasting of an existing possibly sensitive search is difficult, or the final state has evaded attention until now. We use the \texttt{CheckMATE\;2}\,\cite{Dercks:2016npn,Cacciari:2005hq,Cacciari:2008gp,Cacciari:2011ma,Read:2002hq,checkmate} 
framework for testing our benchmarks against the current LHC bounds. We also implement the recasting of a few searches, which we find important for our UDD benchmarks and which were absent in the \texttt{CheckMATE\;2} framework, and discuss their importance in improving the bounds.  

The rest of the paper is organised as follows: in Section\,\ref{sec:framework}, we briefly revisit our previous framework from Ref.\,\cite{Dreiner_2023}. In Section \ref{sec:UDD}, we apply this framework to the case of UDD couplings, beginning with a discussion of the benchmark scenarios in Section \ref{ssec:benchmarks}, followed by examining the relevant analyses for these benchmarks in Section \ref{ssec:analyses}. We also discuss the searches that we implement within \texttt{CheckMATE\;2} in Sections \ref{sssec:ATLAS_multijet} and \ref{sssec:CMS_leptons8tev}. We describe our analysis setup in Section\,\ref{ssec:setup}. We present our results for both the direct and indirect production of the LSPs in Section\,\ref{sec:results}. Finally, we conclude in Section\,\ref{sec:concl}. 

\section{Revisiting the Framework}
\label{sec:framework}

In this section, we review the framework of Ref.\,\cite{Dreiner_2023} to remind the reader of the conventions used and assumptions made in this study. The most general and renormalisable MSSM superpotential is as follows
\cite{Allanach:2003eb, Dreiner:2023yus}:
\begin{eqnarray}
    W &= W_{\rm RPC} + W_{\rm RPV}, \nonumber \\
    W_{\rm RPV} &= W_{\rm LNV} + W_{\rm BNV},
    \label{eq:superpotential}
\end{eqnarray}
where $W_{\rm RPC}$ contains the $R$-parity conserving terms. 
The second part, $W_{\rm RPV}$, consists of the renormalizable terms violating
$R$-parity, which are further divided into lepton number violating (LNV) and baryon number violating (BNV) terms: 
\begin{eqnarray}
    W_{\rm LNV} = \frac{1}{2}\lambda_{ijk}L_{i}L_{j}\bar{E}_k + \lambda'_{ijk}L_{i}Q_{j}\bar{D}_k& + \kappa_{i}H_{u}L_{i}\,,  \nonumber \\
    W_{\rm BNV} = \frac{1}{2}\lambda''_{ijk}\bar{U}_{i}\bar{D}_{j}\bar{D}_k\,,&
    \label{eq:RPV_terms}
\end{eqnarray}
where $L$, $Q$, and $H_u$ are the MSSM lepton, quark, and the up-type Higgs $SU(2)_L$-doublet chiral 
superfields, respectively, while $\bar{E}$ and $\bar{U}$/$\bar{D}$ are the lepton and quark singlet chiral superfields. The $\lam$'s and $\kappa$'s are, respectively, the trilinear and bilinear couplings and the indices $i,j,k$ run over the three generations. 
We have suppressed the gauge indices. As mentioned in Section \ref{sec:intro}, we focus on the UDD operators in this study, which form the $W_{\rm BNV}$. 
For an easier classification of final states in various scenarios, we introduce the convention used in the present work for the various particles in Table\,\ref{tab:conv}.
\begin{table}[hbt!]
    \centering
    \begin{tabular}{p{2.5cm} p{6cm}}
    \hline\hline
    Symbol     &  Particles\\
    \hline\hline
    $\ell$ & $e/\mu$ \\
    $L$ & $e/\mu/\tau$ \\
    $j_l$ & $u/d/c/s$ jets\\
    $j$ & $j_l$/$b$ jet/decay products of $t$ quark\\
    $V$     & $W/Z/h$\\
    $\tilde{e}$ & $\tilde{e}_L/\tilde{e}_R$\\
    $\tilde{\mu}$ & $\tilde{\mu}_L/\tilde{\mu}_R$\\
    $\tilde{\tau}$ & $\tilde{\tau}_L/\tilde{\tau}_R$\\
    $\tilde{q}_L$ & $\tilde{u}_L/\tilde{d}_L/\tilde{c}_L/\tilde{s}_L$\\
    $\tilde{q}_{3}$ & $\tilde{b}_L/\tilde{t}_L$ \\
    $\tilde{u}$ & $\tilde{u}_R/\tilde{c}_R$ \\
    $\tilde{d}$ & $\tilde{d}_R/\tilde{s}_R$ \\
    $\tilde{W}$ & Winos ($\tilde{W}^0$/$\tilde{W}^\pm$)\\
    $\tilde{H}$ & Higgsinos ($\tilde{H}^0$/$\tilde{H}^\pm$)\\
    \hline\hline
    \end{tabular}
    \caption{Convention for the various SM final states and SUSY particles. For the particles not present in this list, we use the standard convention.}
    \label{tab:conv}
\end{table}

Recall, the LSP is not stable in the RPV-MSSM, and 
thus not a viable dark matter candidate. It is not 
restricted to being electrically neutral or a color singlet. We shall consider the following possible LSPs:
\begin{equation}
    \mathrm{LSP} \in \{\tilde{g},\tilde{q}_L,\tilde{q}_3,\tilde{u},\tilde{d},\tilde{t}_R,\tilde{b}_R,\tilde{\chi}_1^0,\tilde{\chi}^\pm,\tilde{e},\tilde{\mu},\tilde{\tau},\tilde{\nu}\}\,.
\end{equation}

The UDD couplings considered 
here lie in the intermediate range: small enough that the production of the LSP at the colliders is unaffected, but large enough that the LSP decays promptly.
We, therefore, consider that the pair production of SUSY particles and their subsequent decay to the LSP, if the produced particle is not the LSP, all proceed via the RPC-MSSM gauge couplings. The LSP decays via the RPV UDD couplings, and all of the decays involved (including the cascade decays to the LSP) are considered to be prompt. 
The relevant UDD coupling for each benchmark
roughly lies in the range\,\footnote{The exact values of the range depend on the spectrum details and the nature of the UDD coupling involved.},
\begin{eqnarray}
    \sqrt{\frac{(\beta\gamma)10^{-12} {\rm\,GeV}}{m_{\rm LSP}}}\lesssim \lambda'' \ll g\,,
    \label{eq:range}
\end{eqnarray}
where $m_{\rm LSP}, \beta$ and $\gamma$ are respectively the mass, speed and Lorentz 
boost factor of the LSP in a given process.
The LSP decays via the UDD coupling, $\lambda''$, and $g$ denotes a gauge coupling. 
The lower bound is estimated assuming a two-body decay of the LSP having a decay length of 1\,cm in the lab frame.
Smaller values of UDD couplings would lead to interesting phenomenology with displaced signatures from the LSP decay, such as displaced or delayed jets\,\cite{ATLAS:2023oti,CMS:2020iwv,Bhattacherjee:2023kxw}. 
These signatures are orthogonal to the ones we consider in the present work.
We, therefore, keep a systematic and detailed study of the long-lived LSP region of the parameter space for a future study.

We study two production modes of the LSP: direct, and via the cascade decay of another sparticle. 
The final state is, therefore, determined by the produced sparticles, including the cascade to the LSP, the nature of the LSP, and the UDD operator responsible for the LSP decay. 
This can symbolically be written as,
\begin{eqnarray}
    {\rm Final\,state} \sim ({\rm Produced\,sparticle}) \otimes ({\rm LSP}) \otimes ({\rm UDD\,operator})\,.
\end{eqnarray}
Each of them can have the following possibilities:
\begin{itemize}
\item At the LHC, the produced sparticle pairs can be (in decreasing order of production cross section for a fixed mass): $\tilde{g}\tilde{g}$, $\tilde{g}\tilde{q}$/$\tilde{g}\tilde{u}$/$\tilde{g}\tilde{d}$, $\tilde{q}\tilde{q}$/$\tilde{q}_{3}\tilde{q}_{3}$/$\tilde{q}\tilde{u}$, $\tilde{\ell}\tilde{\ell}$/$\tilde{\tau}_L\tilde{\tau}_L$/$\tilde{\ell}\tilde{\nu}$, $\tilde{H}\tilde{H}$, $\tilde{W}\tilde{W}$, $\tilde{B}\tilde{B}$.


\item In the RPV-MSSM, all sparticles can be possible LSPs, leading to a wide variety of final states. We study the status of all possible LSPs --- gluinos, squarks, electroweakinos and sleptons.  

\item Lastly, in Eq.\,(\ref{eq:RPV_terms}), the UDD couplings, $\lam''_{ijk}$, are antisymmetric in $j$ and $k$. Therefore, we 
have 9 different UDD couplings (requiring $j<k$). Since the 
signature for the first two generations of quarks is the same at colliders, the set of UDD couplings having unique collider signatures can be further reduced to four:
\begin{enumerate}
    \item \textbf{UDD} ($i,j,k \in \{1,2\}$) : $\mathbf{\lambda''_{112}}$, $\lambda''_{212}$\,,
    \item {\bf UDD$_3$} ($i,j \in \{1,2\}$, $k=3$) : $\mathbf{\lambda''_{113}}$, $\lambda''_{123}$, $\lambda''_{213}$, $\lambda''_{223}$\,,
    \item {\bf U$_3$DD} ($i=3$, $j,k \in \{1,2\}$) : $\mathbf{\lambda''_{312}}$\,,
    \item {\bf U$_3$DD$_3$} ($i=3$, $j \in \{1,2\}$, $k=3$) : $\mathbf{\lambda''_{313}}$, $\lambda''_{323}$\,,
\end{enumerate}
where the first coupling in each set shown in {\it bold} represents the benchmark coupling from each set that we use for our numerical study.
\end{itemize}

For each LSP, we consider the relevant production modes at the LHC, and study its decay via the four benchmark UDD couplings.
We only keep the masses of the LSP and the produced sparticles within the kinematic reach of the LHC and decouple the rest of the spectrum. 

Note that the renormalization group (RG) evolution of a single UDD coupling at high energy might generate non-zero values for other UDD couplings at lower energies\,\cite{Barbier:2004ez}.
The strongest bound on the product of two RPV UDD couplings comes from the contribution of $K-\bar{K}$ mixing to the $K_L-K_S$ mass difference.
It constrains the product $|\lambda''_{313}\lambda''_{323}| \lesssim 4.8 \times 10^{-4}$ at the weak scale\,\cite{Allanach:1999ic}.
In our framework, both these couplings lead to the same final states in colliders.
If we start with the $\lambda''_{313}$ coupling at the LHC energy scale, the strongest bound on this coupling would arise in the scenario where it generates the $\lambda''_{323}$ coupling with a comparable strength at the weak scale\,\footnote{Note that the RG generated couplings would be typically much smaller than the original coupling as the generation is loop suppressed.}.
Even then it would constrain each of these couplings at $\lambda''_{313}\sim\lambda''_{323}\lesssim 2\times 10^{-2}$.
Our analysis assumes that the UDD couplings are in the range given in Eq.\,(\ref{eq:range}), which for a 1\,TeV LSP has to be greater than $10^{-7}$ and much smaller than $\mathcal{O}(10^{-1})$. 
Since the running of the UDD couplings between the weak scale and the TeV scale is not significant\,\cite{Barger:1995qe}, the above constraint is satisfied within our framework.


\section{Application of the Framework to UDD Couplings}
\label{sec:UDD}

In this section, we first discuss the various benchmark scenarios for each possible LSP and the corresponding final states at the LHC. Depending on these final states, we then discuss the relevant analyses and their availability within the recasting framework, \texttt{CheckMATE\;2}.

\subsection{Benchmark Scenarios}
\label{ssec:benchmarks}

For each possible LSP, we can have different relevant production modes at the LHC leading to different final 
states depending also on the UDD coupling. We discuss 
below the possible production modes and final states for each of the LSPs in the RPV-MSSM for baryon number violating couplings.


\subsubsection*{Gluino LSP}

For a gluino LSP, the dominant production channel at the LHC would be their direct pair production, due
to its high cross section. In the corresponding benchmarks, all the other sparticles 
are decoupled\,\footnote{Note that we still need a squark light enough for the gluino to decay promptly. In this context, ``decoupled'' implies that the sparticle is heavy enough to be beyond the kinematic reach of the collider.
With a related decay width expression taken from Ref.\,\cite{Haber:1984rc}, we find that for a gluino mass of 1-2\,TeV, a squark mass as high as 10\,TeV still leads to the gluino decay length, $c\tau\ll 1$\,mm, for $\lambda''\sim 10^{-7}$.}
from the spectrum.
A gluino LSP cannot decay directly via UDD couplings, it first decays to 
a right-handed off-shell squark and a quark, which 
involves a gauge coupling, and then the squark decays to two quarks in the presence of the UDD coupling. One can find the exact decay chain for this decay and all others that we mention in this study with the RPV \texttt{Python} library \texttt{abc-rpv}\,\cite{Dreiner_2023,abc-rpv}. Table\,\,\ref{tab:gluino_UDD} shows the possible final states for the four benchmark UDD couplings discussed in the previous section for a single gluino LSP. 

\begin{table}[hbt!]
        \centering
        \resizebox{0.65\textwidth}{!}{
        \centering
        \begin{tabular}{c c c c c}
        \hline\hline
        LSP  & Coupling  &   LSP Decay  &  Benchmark Label\\
        \hline\hline
        \multirow{4}{*}{$\tilde{g}$}& $\lambda_{112}''$  &  $3j_l$       &  $D_{\tilde{g}}^{uds}$\\
                                    & $\lambda_{113}''$  &  $2j_l+1b$    &  $D_{\tilde{g}}^{udb}$\\
                                    & $\lambda_{312}''$  &  $2j_l+1t$    &  $D_{\tilde{g}}^{tds}$\\
                                    & $\lambda_{313}''$  &  $1j_l+1b+1t$ &  $D_{\tilde{g}}^{tdb}$\\
        \hline\hline                                                       
        \end{tabular}
        }
        \caption{Details of the gluino LSP benchmarks: the first column depicts the LSP; 
        the RPV coupling assumed to be non-zero is shown in the second column; 
        the third column represents the final state from the individual LSP decay; 
        and the last column shows the notation we use for labeling the benchmark scenario when the LSP is directly produced, following similar conventions in Ref.\,\cite{Dreiner_2023}. }
        \label{tab:gluino_UDD}
    \end{table}

\subsubsection*{Squark LSPs}

For squark LSPs, we consider the direct pair production of the squarks at the LHC, where all the other sparticles are decoupled (see footnote 3) from the spectrum.
In addition, we consider the gluino-squark model, where we have  $\tilde{g}\tilde{g}$ pair production and $\tilde{g}\tilde{q}$ associated production at the LHC. In this latter scenario, the second squark 
LSP is produced 
from the gluino decay. Note that the associated production is only possible for the first two generation squarks, assuming a 4-flavor parton 
distribution function (PDF) scheme. 

\begin{table}[hbt!]
    \centering
    \resizebox{0.7\textwidth}{!}{
    \begin{tabular}{c | c c c c}
    \hline\hline
    \multirow{2}{*}{LSP} & \multicolumn{4}{c}{Coupling} \\ \cline{2-5}
         & $\lambda''_{112}$ & $\lambda''_{113}$ & $\lambda''_{312}$ & $\lambda''_{313}$ \\
    \hline\hline
    $\tilde{u}_R$ &  \textcolor{greeen}{Direct} &  \textcolor{greeen}{Direct} &Cascade &Cascade \\
    $\tilde{d}_R$ &  \textcolor{greeen}{Direct} &  \textcolor{greeen}{Direct} &  \textcolor{greeen}{Direct} &  \textcolor{greeen}{Direct} \\
    $\tilde{t}_R$ &Cascade &Cascade &  \textcolor{greeen}{Direct} &  \textcolor{greeen}{Direct} \\
    $\tilde{b}_R$ &Cascade &  \textcolor{greeen}{Direct} &Cascade &  \textcolor{greeen}{Direct} \\
    $\tilde{q}_L/\tilde{b}_L/\tilde{t}_L$ &Cascade &Cascade &Cascade &Cascade \\
    \hline\hline
    \end{tabular}
    }
    \caption{Direct or cascade decays of the various squark LSPs for the four benchmark UDD couplings.}
    \label{tab:squark_decays}
\end{table}

\begin{table}[hbt!]
    \centering
    \resizebox{0.725\textwidth}{!}{
    \begin{tabular}{c c c c}
    \hline\hline
    LSP  &   Coupling  &  LSP Decay & Benchmark Labels  \\
    \hline\hline
    \multirow{4}{*}{$\tilde{u}_R$} &  $\lambda_{112}''$ & \textcolor{greeen}{$2j_l$}    & $D_{\tilde{u}}^{uds}$             \\
                                   &  $\lambda_{113}''$ & \textcolor{greeen}{$1j_l+1b$} & $D_{\tilde{u}}^{udb}$             \\
                                   &  $\lambda_{312}''$ & $3j_l+1t$                     & $D_{\tilde{u}}^{tds}$             \\
                                   &  $\lambda_{313}''$ & $2j_l+1t+1b$                  & $D_{\tilde{u}}^{tdb}$             \\ \hline
    \multirow{4}{*}{$\tilde{d}_R$} &  $\lambda_{112}''$ & \textcolor{greeen}{$2j_l$}    & $D_{\tilde{d}}^{uds}$             \\
                                   &  $\lambda_{113}''$ & \textcolor{greeen}{$1j_l+1b$} & $D_{\tilde{d}}^{udb}$             \\
                                   &  $\lambda_{312}''$ & \textcolor{greeen}{$1j_l+1t$} & $D_{\tilde{d}}^{tds}$             \\
                                   &  $\lambda_{313}''$ & \textcolor{greeen}{$1t+1b$}   & $D_{\tilde{d}}^{tdb}$             \\ \hline
    \multirow{4}{*}{$\tilde{t}_R$} &  $\lambda_{112}''$ & $3j_l+1t$                     & $D_{\tilde{t}}^{uds}$             \\
                                   &  $\lambda_{113}''$ & $2j_l+1t+1b$                  & $D_{\tilde{t}}^{udb}$             \\
                                   &  $\lambda_{312}''$ & \textcolor{greeen}{$2j_l$}    & $D_{\tilde{t}_R}^{tds}$           \\
                                   &  $\lambda_{313}''$ & \textcolor{greeen}{$1j_l+1b$} & $D_{\tilde{t}_R}^{tdb}$           \\ \hline
    \multirow{4}{*}{$\tilde{b}_R$} &  $\lambda_{112}''$ & $3j_l+1b$                     & $D_{\tilde{b}}^{uds}$             \\
                                   &  $\lambda_{113}''$ & \textcolor{greeen}{$2j_l$}    & $D_{\tilde{b}_R}^{udb}$           \\
                                   &  $\lambda_{312}''$ & $2j_l+1t+1b$                  & $D_{\tilde{b}}^{tds}$             \\
                                   &  $\lambda_{313}''$ & \textcolor{greeen}{$1j_l+1t$} & $D_{\tilde{b}_R}^{tdb}$           \\ \hline
    \multirow{4}{*}{$\tilde{q}_L$} &  $\lambda_{112}''$ & $4j_l$                        & $D_{\tilde{q}}^{uds}$             \\
                                   &  $\lambda_{113}''$ & $3j_l+1b$                     & $D_{\tilde{q}}^{udb}$             \\
                                   &  $\lambda_{312}''$ & $3j_l+1t$                     & $D_{\tilde{q}}^{tds}$             \\ 
                                   &  $\lambda_{313}''$ & $2j_l+1t+1b$                  & $D_{\tilde{q}}^{tdb}$             \\ \hline
    \multirow{4}{*}{$\tilde{t}_L$} &  $\lambda_{112}''$ & $3j_l+1t$                     & $D_{\tilde{t}}^{uds}$             \\
                                   &  $\lambda_{113}''$ & $2j_l+1t+1b$                  & $D_{\tilde{t}}^{udb}$             \\
                                   &  $\lambda_{312}''$ & $2j_l+2t / 2j_l+2b$           & $D_{\tilde{t}_L}^{tds}$           \\
                                   &  $\lambda_{313}''$ & $1j_l+2t+1b / 1j_l+3b$        & $D_{\tilde{t}_L}^{tdb}$           \\ \hline
    \multirow{4}{*}{$\tilde{b}_L$} &  $\lambda_{112}''$ & $3j_l+1b$                     & $D_{\tilde{b}}^{uds}$             \\
                                   &  $\lambda_{113}''$ & $2j_l+2b / 2j_l+2t$           & $D_{\tilde{b}_L}^{udb}$           \\
                                   &  $\lambda_{312}''$ & $2j_l+1t+1b$                  & $D_{\tilde{b}}^{tds}$             \\
                                   &  $\lambda_{313}''$ & $1j_l+1t+2b / 1j_l+3t$        & $D_{\tilde{b}_L}^{tdb}$           \\ 
    \hline\hline                            
    \end{tabular}
    }
    \caption{Details of the squark LSP benchmarks with columns as in Table\,\,\ref{tab:gluino_UDD}.
    We show the final states from the direct decay (one step cascade decay) in {\it \textcolor{greeen}{blue}} ({\it black}).}

    \label{tab:squark_UDD}    
\end{table}

Depending on the UDD coupling and the squark flavor, the left- and right-handed squark decays lead to different final states.
For example, a right-handed up-type squark decays directly to two quarks via the $\lam''_{112}$ or $\lam''_{113}$ couplings.
However, in case of the couplings $\lambda''_{312}$ or $\lambda''_{313}$, $\tilde{u}_R$ has to decay via two cascades (say, $\tilde{u}_R\rightarrow u_R \tilde{g}$, $\tilde{g}\rightarrow d_R \tilde{d}_R$) before the decay via the UDD operator.
This increases the multiplicity of the final state and therefore affects the bounds on the different squark LSPs.
In Table\,\ref{tab:squark_decays}, we show whether the squark LSP has a ``direct'' or ``cascade'' decay for each of the four benchmark UDD couplings.

Table\,\ref{tab:squark_UDD} shows the possible final states for the four benchmark UDD couplings for the different squark LSPs.
We denote the direct decays of the squarks in {\it \textcolor{greeen}{blue}}, while the one step cascade decays are denoted in {\it black}.
Note that the heavy flavor left-handed squarks have two possible cascade decays: one via a gluino or bino, and the second via a charged higgsino. 
When we consider the production mode $\tilde{g}\tilde{g}$ ($\tilde{g}\tilde{q}$), there will be two (one) extra jets in each of the final states mentioned in Table\,\ref{tab:squark_UDD}. 

\medskip

\subsubsection*{Electroweakino LSPs}

In the electroweakino sector, we first consider the bino-dominated LSP, $\tilde{B}$. These 
have a similar cascade decay as the gluino LSPs in the presence of the UDD operators, as 
shown in Table\,\ref{tab:Bino_UDD}. However, the pair production cross section for 
$\tilde{B}$ is very small at the LHC. We, therefore, only consider its production via the decay of some other sparticle in the spectrum, which has a higher 
production cross section, like gluinos and squarks. When we consider the production mode 
$\tilde{g}\tilde{g}$, there will be four extra jets in each of the final states mentioned 
in Table\,\ref{tab:Bino_UDD}. In the case of squarks, we will have two extra quarks in each of 
the final states of 
Table\,\ref{tab:Bino_UDD}. The flavor of each quark is the same as that of the produced 
squark, which decays to the $\tilde{B}$ LSP.
\begin{table}[hbt!]
    \centering
    \resizebox{0.65\textwidth}{!}{
    \centering
    \begin{tabular}{c c c c}
    \hline\hline
    LSP  &  Coupling  &   LSP Decay  &  Benchmark Label\\
    \hline\hline
    \multirow{4}{*}{$\tilde{B}$}  & $\lambda_{112}''$ & $3j_l$       &  $D_{\tilde{B}}^{uds}$\\ 
                                  & $\lambda_{113}''$ & $2j_l+1b$    &  $D_{\tilde{B}}^{udb}$\\ 
                                  & $\lambda_{312}''$ & $2j_l+1t$    &  $D_{\tilde{B}}^{tds}$\\ 
                                  & $\lambda_{313}''$ & $1j_l+1b+1t$ &  $D_{\tilde{B}}^{tdb}$\\ 
    \hline\hline                                                     
\end{tabular}
    }
    \caption{Details of the bino LSP benchmarks with columns as in Table\,\,\ref{tab:gluino_UDD}.}
    \label{tab:Bino_UDD}
\end{table}

\begin{table}[hbt!]
    \resizebox{\textwidth}{!}{
    \centering
    \begin{tabular}{c c c c}
    \hline\hline
    LSP  &  Coupling  &   LSP Decay  &  Label\\
    \hline\hline

    \multirow{4}{*}{$\tilde{W}^0$}  & $\lambda_{112}''$ &  \textcolor{brown}{$3j_l +$ ($2j_l/2b/2t/2L/2V/$MET)}     &  $D_{\tilde{W}}^{uds}$\\
                                  &  $\lambda_{113}''$ & \textcolor{newpurple}{$2j_l + 1b + 1V$ \,\,\, $2j_l + 1t + 1V$} \,\,\, \textcolor{brown}{$2j_l + 1b +$ ($2j_l/2b/2t/2L/2V/$MET)}     &  $D_{\tilde{W}}^{udb}$\\
                                  &  $\lambda_{312}''$ & \textcolor{newpurple}{$2j_l + 1t + 1V$ \,\,\, $2j_l + 1b + 1V$} \,\,\, \textcolor{brown}{$2j_l + 1t +$ ($2j_l/2b/2t/2L/2V/$MET)}        &  $D_{\tilde{W}}^{tds}$\\
                                  &  $\lambda_{313}''$ &  \textcolor{newpurple}{$1j_l + 1t + 1b + 1V$ \,\,\, $1j_l + 2b + 1V$ \,\,\, $1j_l + 2t + 1V$} \,\,\, \textcolor{brown}{$1j_l + 1t + 1b +$ ($2j_l/2b/2t/2L/2V/$MET)}  &  $D_{\tilde{W}}^{tdb}$\\\hline
    \multirow{4}{*}{$\tilde{W}^\pm$}  & $\lambda_{112}''$ & \textcolor{brown}{$3j_l + $ ($2j_l/2V$)} \,\,\, \textcolor{brown}{$3j_l + 1t + 1b$} \,\,\, \textcolor{brown}{$3j_l + 1L +$MET}  &  $D_{\tilde{W}}^{uds}$\\
                                  &  $\lambda_{113}''$ &  \textcolor{newpurple}{$2j_l + 1t + 1V$ \,\,\, $2j_l + 1b + 1V$} \,\,\, \textcolor{brown}{$2j_l + 1t + 2b$} \,\,\, \textcolor{brown}{$2j_l + 1b +$ ($2j_l/2V$)} \,\,\, \textcolor{brown}{$2j_l + 1b + 1L +$MET}  &  $D_{\tilde{W}}^{udb}$\\
                                  &  $\lambda_{312}''$ & \textcolor{newpurple}{$2j_l + 1b + 1V$ \,\,\, $2j_l + 1t + 1V$} \,\,\, \textcolor{brown}{$2j_l + 2t + 1b$} \,\,\, \textcolor{brown}{$2j_l + 1t +$ ($2j_l/2V$)} \,\,\, \textcolor{brown}{$2j_l + 1t + 1L +$MET}    &  $D_{\tilde{W}}^{tds}$\\
                                  & \multirow{2}{*}{ $\lambda_{313}''$} & \textcolor{newpurple}{$1j_l + 1t + 1b + 1V$ \,\,\, $1j_l + 2b + 1V$ \,\,\, $1j_l + 2t + 1V$} &  \multirow{2}{*}{$D_{\tilde{W}}^{tdb}$}\\
                                  &                    & \textcolor{brown}{$1j_l + 1t + 1b + $ ($2j_l/2V$)} \,\,\, \textcolor{brown}{$1j_l + 2t + 2b$} \,\,\, \textcolor{brown}{$1j_l + 1t + 1b + 1L +$MET} & \\
    \hline\hline                                                       
    \end{tabular}
     }
    \caption{Details of the wino LSP benchmarks with columns as in Table\,\ref{tab:gluino_UDD}. We show the final states from a two (three) step cascade decay in {\it \textcolor{newpurple}{purple}} ({\it \textcolor{brown}{brown}}).}
    \label{tab:W_UDD}
\end{table}

The wino-dominated LSPs, $\tilde{W}$, can be directly pair-produced at the LHC ($\tilde
{\chi}_1^0\tilde{\chi}_1^\pm$ or $\tilde{\chi}_1^\pm\tilde{\chi}_1^\pm$) when the rest 
of the spectrum is decoupled, or they can also come from the decay of the strong sector 
sparticles, which have higher cross sections. The wino-like LSPs decay via UDD couplings 
through longer cascades, since they only couple to left-handed 
sparticles, while the UDD operators involve only the right-handed fields. A typical 
cascade for the wino-like neutralino would be (with $^\star$ denoting off-shell states):
$\tilde{\chi}_1^0\to j \tilde{q}^\star_L \rightarrow jj \tilde{g}^\star\rightarrow 
jjj \tilde{u}^\star_R \rightarrow jjjjj$ for the $\lambda''_{112}$ coupling, which is a 
three step cascade before the final UDD decay. The intermediate sparticle can also be a bino 
or a higgsino instead of a gluino. For UDD operators involving heavy flavors, 
\textit{i.e.} $\lam''_{113}, 
\lam''_{312}, \lam''_{313}$, we can have a shorter cascade due to the large Yukawa 
couplings of the heavy flavor quarks, where the higgsinos in the intermediate steps can 
mix the left and the right-handed fields. We list all the possible final states for the 
wino LSPs in Table\,\ref{tab:W_UDD} for the four benchmark UDD couplings. The final states 
shown in {\it \textcolor{newpurple}{purple}} are from a two step cascade decay, while the 
ones shown in {\it \textcolor{brown}{brown}} result from a three step cascade decay.




The higgsino-dominated LSPs, $\tilde{H}$, can be directly pair-produced at the LHC
($\tilde{\chi}_{1/2}^0\tilde{\chi}_1^\pm$ or $\tilde{\chi}_1^\pm\tilde{\chi}_1^\pm$) 
when the rest of the spectrum is decoupled, or they can also come from the decay of 
gluinos and heavy flavor squarks. The length of the cascades for higgsino-like LSP 
decays via UDD couplings are shorter than that for the wino-like LSPs. The possible 
final states for the higgsino LSPs are listed in Table\,\ref{tab:H_UDD}
for the four benchmark UDD couplings. The final states shown in {\it black} are from 
a one step cascade decay, while the ones shown in {\it \textcolor{newpurple}{purple}}
result from a two step cascade decay.



\begin{table}[hbt!]
    \centering
    \resizebox{0.80\textwidth}{!}{
    \centering
    \begin{tabular}{c c c c}
    \hline\hline
    LSP  &  Coupling  &   LSP Decay  &  Benchmark Label\\
    \hline\hline
    \multirow{4}{*}{$\tilde{H}^0$}  &  $\lambda_{112}''$ &  \textcolor{newpurple}{$3j_l + 1V $}                                      &  $D_{\tilde{H}}^{uds}$\\
                                  &  $\lambda_{113}''$ &  $2j_l + 1b$ \,\,\, \textcolor{newpurple}{$2j_l + 1b + 1V $}   &  $D_{\tilde{H}}^{udb}$\\
                                  &  $\lambda_{312}''$ &  $2j_l + 1t$ \,\,\, \textcolor{newpurple}{$2j_l + 1t + 1V $} &  $D_{\tilde{H}}^{tds}$\\
                                  &  $\lambda_{313}''$ &  $1j_l + 1b + 1t$ \,\,\, \textcolor{newpurple}{$1j_l + 1b + 1t + 1V$} &  $D_{\tilde{H}}^{tdb}$\\ \hline
    \multirow{4}{*}{$\tilde{H}^\pm$}  & $\lambda_{112}''$ &  \textcolor{newpurple}{$3j_l + 1V $}                                      &  $D_{\tilde{H}}^{uds}$\\
                                  &  $\lambda_{113}''$ &  $2j_l + 1t$ \,\,\, \textcolor{newpurple}{$2j_l + 1b + 1V $}  &  $D_{\tilde{H}}^{udb}$\\
                                  &  $\lambda_{312}''$ & $2j_l + 1b$ \,\,\, \textcolor{newpurple}{$2j_l + 1t + 1V $}  &  $D_{\tilde{H}}^{tds}$\\
                                  &  $\lambda_{313}''$ & $1j_l + 2b$ \,\,\, $1j_l + 2t$ \,\,\, \textcolor{newpurple}{$1j_l + 1b + 1t + 1V$}  &  $D_{\tilde{H}}^{tdb}$\\                          
    \hline\hline                                                       
    \end{tabular}
    }
    \caption{Details of the higgsino LSP benchmarks with columns as in Table\,\ref{tab:gluino_UDD}. We show the final states from a one (two) step cascade decay in {\it \textcolor{black}{black}} ({\it \textcolor{newpurple}{purple}}).}
    \label{tab:H_UDD}
\end{table}

\subsubsection*{Slepton LSPs}

Slepton LSPs can be directly pair-produced at the LHC, or they can come from the decay of gluinos or winos. The sleptons decay via a two step cascade with a bino in the intermediate step, while they have longer cascades when they decay via wino-like intermediate sparticles. We show the possible final states for the sleptons having a two step (three step) cascade decay in
Table\,\ref{tab:nv2slep_UDD} (Table\,\ref{tab:nv3slep_UDD} in Appendix\,\ref{app:plots}) for the four benchmark UDD couplings. 

\begin{table}[hbt!]
    \centering
    \resizebox{0.65\textwidth}{!}{
    \centering
    \begin{tabular}{c c c c}
    \hline\hline
    LSP  &  Coupling  &   LSP Decay  &  Benchmark Label\\
    \hline\hline
    \multirow{4}{*}{$\tilde{e}/\tilde{\mu}/\tilde{\tau}$}  &  $\lambda_{112}''$  &  \textcolor{newpurple}{$3j_l   + 1e/\mu/\tau$}                                          &  $D_{\tilde{e}}^{uds}$\\
                                  &  $\lambda_{113}''$  &  \textcolor{newpurple}{$2j_l+1b+ 1e/\mu/\tau$}                                      &  $D_{\tilde{e}}^{udb}$\\
                                  &  $\lambda_{312}''$  & \textcolor{newpurple}{$2j_l+1t+ 1e/\mu/\tau$} &  $D_{\tilde{e}}^{tds}$\\
                                  &  $\lambda_{313}''$  & \textcolor{newpurple}{$1j_l+1b+1t+ 1e/\mu/\tau$} &  $D_{\tilde{e}}^{tdb}$\\\hline
    \multirow{4}{*}{$\tilde{\nu}_e/\tilde{\nu}_\mu/\tilde{\nu}_\tau$} &  $\lambda_{112}''$  &  \textcolor{newpurple}{$3j_l+$MET}                                       &  $D_{\tilde{\nu}}^{uds}$\\
                                  &  $\lambda_{113}''$  &  \textcolor{newpurple}{$2j_l+1b+$MET  }                                       &  $D_{\tilde{\nu}}^{udb}$\\
                                  &  $\lambda_{312}''$  & \textcolor{newpurple}{$2j_l+1t+$MET} &  $D_{\tilde{\nu}}^{tds}$\\
                                  &  $\lambda_{313}''$  & \textcolor{newpurple}{$1j_l+1b+1t+$MET} &  $D_{\tilde{\nu}}^{tdb}$\\                              
    \hline\hline                                                       
    \end{tabular}
    }
    \caption{Details of the slepton LSP benchmarks when the decay happens via a two step cascade (shown in {\it \textcolor{newpurple}{purple}}) involving an off-shell bino, with columns as in Table\,\ref{tab:gluino_UDD}.}
    \label{tab:nv2slep_UDD}
\end{table}

\subsection{Analyses Relevant for the Benchmarks}
\label{ssec:analyses}

In Ref.\,\cite{Dreiner_2023}, the signatures of LSP decays via UDD couplings were broadly classified into the following:

\begin{enumerate}
    \item $4j$, 

    \item $2j_l + 4j$, 

    \item $2j_l + 6j$, 

    \item $1L + 2j_l + 4j + E_T^{\rm miss}$, 

    \item $2L + 2j_l + 4j$, 

\end{enumerate}
where $L$, $j$, $j_l$ are defined in Table\,\ref{tab:conv}. The 
CMS and ATLAS searches that might be sensitive to these final states were given in Ref.\,\cite{Dreiner_2023}. We tabulate them in 
Table\,\ref{tab:searches}. 
As we see, only one 
of these searches is implemented in \texttt{CheckMATE\;2}, which limits our knowledge 
of the coverage of UDD couplings when we recast using \texttt{CheckMATE\;2}.

\begin{table}[hbt!]
    \centering
    \resizebox{\textwidth}{!}{
    \begin{tabular}{c c c | c}
    \hline\hline
      \multirow{2}{*}{Final state} & Possible sensitive & Implemented in & \multirow{2}{*}{Comment} \\
       & searches & \texttt{CheckMATE\;2}? &  \\
      \hline\hline
      \multirow{2}{*}{4 jets} &  CMS\,\cite{CMS:2022usq} & \no{No} & \multirow{2}{*}{Used CMS\,\cite{CMS:2022usq} directly}\\
                              & ATLAS\,\cite{ATLAS:2023ssk} & \no{No} & \\

      \hline
      \multirow{1}{*}{6 jets} & ATLAS\,\cite{ATLAS:2024kqk} & \no{No} & Implemented ATLAS\,\cite{ATLAS:2024kqk}\\
      \cline{1-3}
      \multirow{1}{*}{8 jets} & ATLAS\,\cite{ATLAS:2024kqk} & \no{No} & for this work\\
      \hline
      \multirow{5}{*}{1 lepton + $\geq$ 6 jets} & CMS\,\cite{CMS:2020cur} & \no{No} & \multirow{5}{*}{$-$}\\
                              & ATLAS\,\cite{ATLAS:2021fbt} & \yes{Yes} & \\
                              & CMS\,\cite{CMS:2022cpe} & \no{No} & \\
                              & ATLAS\,\cite{OliveiraCorrea:2025dlq} & \no{No} & \\
                              & ATLAS\,\cite{ATLAS:2021upq} & \no{No} & \\
                              
      \hline
      \multirow{4}{*}{2 leptons + 6 jets} & CMS\,\cite{CMS:2020cpy} & \no{No} & \multirow{3}{*}{Implemented CMS\,\cite{CMS:2016ooq} }\\
                              & CMS\,\cite{CMS:2022cpe} & \no{No} & \multirow{3}{*}{for this work}\\
                              & CMS\,\cite{CMS:2016ooq} & \no{No} & \\
                              & ATLAS\,\cite{OliveiraCorrea:2025dlq} & \no{No} & \\
      \hline\hline                        
    \end{tabular}
    }
    \caption{List of relevant analyses for the UDD final states along with their status of implementation in the recasting framework, \texttt{CheckMATE\;2}.}
    \label{tab:searches}
\end{table}

Therefore, we implement some of these searches in order to reduce the gap between
the coverage of the recasting framework available to theorists and the actual 
experimental coverage. In this section, we discuss three different searches 
implemented by us in \texttt{CheckMATE\;2} for improving sensitivity to UDD 
couplings. We picked searches where detailed information regarding the 
experimental input and results was available. We select the ATLAS multijet 
search\,\cite{ATLAS:2024kqk} which has sensitivity to final states involving 
6--8\,jets. The next search we implement is the CMS search with two leptons and 
multiple jets in the final state\,\cite{CMS:2016ooq}. Searches with 4 jets in the 
final state are based on fitting distributions of pairs of jets to identify the 
new physics resonance. Implementing such a search in \texttt{CheckMATE\;2} is 
difficult, and therefore, we use the result from the CMS study\,\cite{CMS:2022usq} 
directly for cases where we have a 4 jets final state. 
In the last column of Table\,\ref{tab:searches} we comment on how we got coverage 
for the five listed classes of final states. Additionally, we implement a leptoquark 
search to explore whether it can enhance sensitivity for LSP decays with leptons in 
the final state. We discuss the three searches implemented in the following subsections.

\subsubsection{Implementation of the ATLAS Multijet Search at 13\,TeV}
\label{sssec:ATLAS_multijet}

UDD signal events in the colored LSP sector produce a large number of jets, see 
Tables\,\ref{tab:gluino_UDD} and \ref{tab:squark_UDD}. Unfortunately, there was no 
multijet search implemented in the default analysis library of our recasting tool 
\texttt{CheckMATE\;2}, as of yet. So we manually implemented the multijet search 
\textbf{atlas\_2401\_16333}\,\cite{ATLAS:2024kqk} into \texttt{CheckMATE\;2} 
following the procedure described in Ref.\,\cite{Kim:2015wza}. The search utilizes 
$\SI{140}{\per\femto\barn}$ of proton–proton collision data at $\sqrt{s} = \SI{13}
{\tera\electronvolt}$, recorded by the ATLAS experiment during Run 2 of the LHC. The 
results were analyzed within the framework of RPV-SUSY models, which involve prompt 
gluino pair production with subsequent decay of a gluino into either three jets, or 
into two jets and a neutralino that promptly decays into three jets. The results of 
this study are thus very well suited for recasting within our generalization of RPV 
signatures.

The details of the event reconstruction can be found in 
Sec.\,4 of Ref.\,\cite{ATLAS:2024kqk}.
Let us note here that the anti-$k_T$ algorithm with a cone size of $R = 0.4$ is used to define analysis-level jets. We then define ``baseline" jets with $p_T > \SI{20}{\giga\electronvolt}$ and $|\eta| < 4.8$, from which we further select ``signal" jets with $p_T > \SI{50}{\giga\electronvolt}$ and $|\eta| < 2.8$.
Jets with $|\eta| < 2.5$ are tagged as $b$-jets by a multivariate algorithm (DL1r)\,\cite{ATLAS:2022qxm,ATLAS:2024kqk} 
with a tagging efficiency of $77\,\%$.
For the analysis, only events that contain at least 4 ``signal" jets and no electron or muon candidates are selected.

Signal and background are well separated in terms of the event shape variable $C$ defined as,
\begin{equation}
C \;\equiv\; 3 (\lam_1 \lam_2 + \lam_1 \lam_3 + \lam_2 \lam_3)\,,
\end{equation}
where the $\lam_i$ are the eigenvalues of the linearized sphericity tensor of 
each event, which contains only jet momenta after the previous selections. The 
event shape variable provides a measure of the isotropy of an event: small values 
indicate collimated events, while large values suggest more spherical, isotropic 
distributions of jet momenta. \textbf{atlas\_2401\_16333} defines in total seven 
signal regions featuring a large number of high-$p_T$ jets as well as a tight 
selection of large $C$. Since we consider the pair production of heavy sparticles with a subsequent UDD cascade decay in our study, we expect a large number of jets 
isotropically distributed in the final state.
This makes \textbf{atlas\_2401\_16333} very well suited to exclude large parts of the RPV-SUSY parameter space, after recasting the results.

\begin{figure}[h!]
    \centering
    \includegraphics[width=0.69\linewidth]{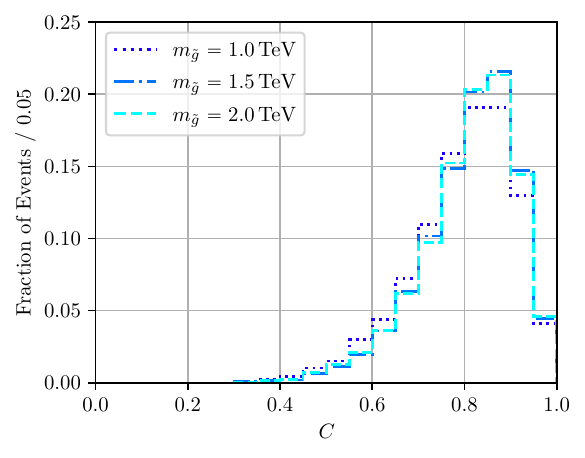}
    \caption{Distribution of the shape variable, $C$, for three different masses of the gluino LSP decaying via the $\lambda''_{112}$ coupling obtained from the \texttt{CheckMATE\;2} implementation of \textbf{atlas\_2401\_16333}. 
    }
    \label{fig:cvalue}
\end{figure}
\begin{figure}[h!]
    \centering
    \includegraphics[width=\textwidth]{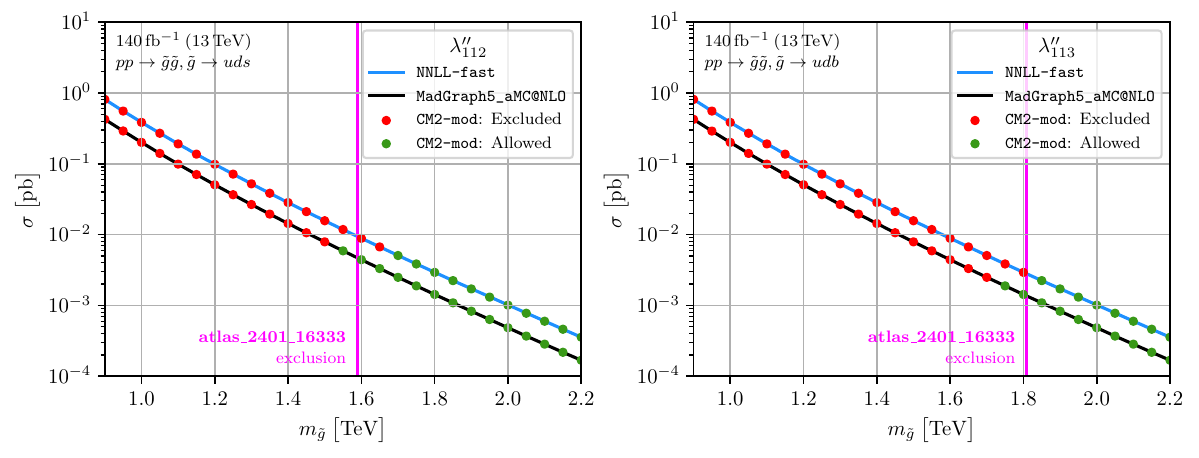}
    \caption{Validation of the implementation of the ATLAS multijet search\,\cite{ATLAS:2024kqk} in \texttt{CheckMATE\;2} for the $\lambda''_{112}$ ({\it left}) and $\lambda''_{113}$ ({\it right}) UDD couplings for the gluino LSP decay with the LO cross sections from \texttt{MadGraph5\_aMC@NLO} and the NNLO+NNLL cross sections from \texttt{NNLL-fast}.}
    \label{fig:atlas_mj_validation}
\end{figure}
To validate our \texttt{CheckMATE\;2} implementation of the search, we ran 
\texttt{CheckMATE\;2} for the same RPV-SUSY scenario and read out the values of the 
event shape variable. 
Their distribution is shown in Fig.\,\ref{fig:cvalue}; within 
statistics it is in good agreement with the distribution obtained by the ATLAS collaboration
(see Fig.\,2(b) in Ref.\,\cite{ATLAS:2024kqk}).
Furthermore, for the $\tilde{g} \rightarrow uds$ and $\tilde{g} \rightarrow udb$ 
RPV decays, we scan through the gluino masses to find the \texttt{CheckMATE\;2} 
exclusion limits. For each case, we do this for both the leading-order (LO) cross 
sections obtained from \texttt{MadGraph5\_aMC@NLO} \cite{Alwall:2014hca} and NNLO+NNLL cross sections from \texttt{NNLL-fast} (Refs.\,\cite{Beenakker:2016lwe, Beenakker:2024jwh}).
We compute the $r$-value for each signal region using \texttt{CheckMATE\;2}, which is defined in Ref.\,\cite{Dercks:2016npn}. A signal benchmark having an $r$-value greater than 1 in any of the signal regions is excluded.
The result is shown in Fig.\,\ref{fig:atlas_mj_validation}.
When using the LO cross section, the obtained limit is always lower than the ATLAS value, as expected.
With \texttt{NNLL-fast}, we reproduce the ATLAS limit within our $\SI{50}{\giga\electronvolt}$ grid in the $\lambda^{\prime\prime}_{113}$ case.
For $\lambda^{\prime\prime}_{112}$, \texttt{CheckMATE\;2} slightly overestimates the exclusion limit.

\subsubsection{Implementation of the CMS two OSSF Leptons along with Jets Search at 8\,TeV}
\label{sssec:CMS_leptons8tev}

We have discussed in Section\,\ref{ssec:benchmarks} that the charged slepton LSP decays always involve a charged lepton in the final state. In the pair production of sleptons, this leads to two opposite-sign same-flavor (OSSF) leptons along with multiple jets. We find that a relevant search for this final state was performed by the CMS collaboration at 8\,TeV LHC in Ref.\,\cite{CMS:2016ooq}. This search is not implemented in \texttt{CheckMATE\;2} yet. We implement it for the present work and discuss the details of the implementation here. 

The CMS search for two OSSF leptons along with multiple jets in the final 
state in Ref.\,\cite{CMS:2016ooq} uses 19.7\,fb$^{-1}$ of data collected 
at a center-of-mass energy of 8\,TeV during the Run 1 of LHC. It focuses 
on the stop search, where the stop decays via $\tilde{t}\rightarrow b\tilde
{\chi}^\pm$, $\tilde{\chi}^\pm\rightarrow l^\pm jj$, where the second decay 
occurs due to the presence of a single RPV LQD coupling, $\lambda'_{ijk}$ 
($i,j,k\leq2$). This final state is similar to the one we have for charged 
slepton decays via UDD couplings in Table\,\ref{tab:nv2slep_UDD}. We closely
follow the event selection criteria described in Ref.\,\cite[Sec.\,3]{CMS:2016ooq}. 
We briefly discuss the details below for completeness and highlight the 
exact settings we use in \texttt{CheckMATE\;2}.

\begin{sloppypar}
While implementing this search in \texttt{CheckMATE\;2}, we use electrons 
reconstructed in the \texttt{electronsLoose} category, while for muons, we use 
the \texttt{muonsCombined} category. Electron (muon) candidates are required 
to have $p_T>50$\,GeV and $|\eta|<2.5$ ($|\eta|<2.1$). Jets are clustered 
using the anti-$k_T$ algorithm with $R=0.4$, and are required to satisfy 
$|\eta|<2.4$. The leading jet must have a transverse momentum of at least 
100\,GeV, the sub-leading jet should have $p_T>50$\,GeV, and the rest of the 
jets should satisfy $p_T>30$\,GeV each. We also use 
$b$-tagging with a tagging efficiency of 70\%. The analysis selects events 
with two oppositely charged electrons or muons and a minimum of five jets, at 
least one of which is $b$ tagged.
\end{sloppypar}

\begin{figure}[hbt!]
    \centering
    \includegraphics[width=\textwidth]{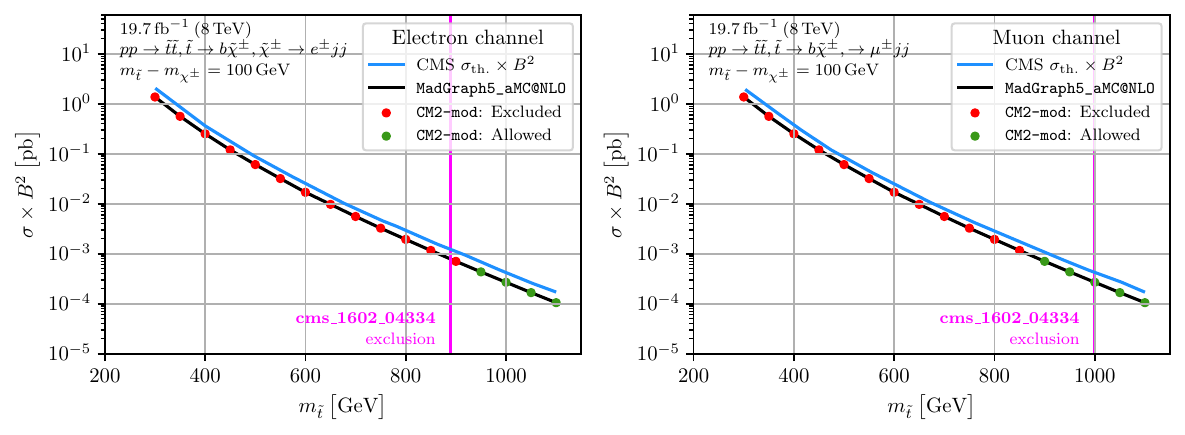}\,
    \caption{Validation of the implementation of the CMS OSSF lepton pair with jets search at 8\,TeV\,\cite{CMS:2016ooq} in \texttt{CheckMATE\;2} for the chargino-mediated stop decay via LQD couplings, with 100\% branching to $e^\pm e^\mp$ ({\it left}) and to $\mu^\pm \mu^\mp$ ({\it right}).}
    \label{fig:val_CMS_lep_8tev}
\end{figure}

In order to suppress the large $t\bar{t}$ background for the case of 
the leptonic decay of the top quark, events with a $E_T^{\rm miss}$ larger than 
100\,GeV are vetoed. To reduce the background from the decays of low-mass resonances and the $Z$ 
boson to leptons, the invariant mass of the OSSF lepton pair must be higher than 
130\,GeV. Depending on the flavor of the leptons, the signal regions are divided 
into two channels: the electron channel and the muon channel. Within each 
channel, the signal regions are further divided based on the number of jets in 
the final state and the minimum value of $S_T$, the scalar sum of the transverse 
momenta of all the jets and leptons in the event. The minimum value of $S_T$ 
required is 
optimised for various stop masses in the analysis for a given number of jets in 
the final state. We use the numbers for the observed data and the expected 
background events as given in Ref.\,\cite[Tables\,3,4]{CMS:2016ooq} to calculate 
the signal significance using \texttt{CheckMATE\;2}. Fig.\,\ref{fig:val_CMS_lep_8tev} 
shows the validation of our implementation of this analysis. We find that for the 
electron channel our implementation matches the CMS result within a step size of 
100\,GeV, however, for the muon channel the bound obtained from our implementation 
differs from the CMS bound by 200\,GeV. Our implementation, 
therefore, underestimates the exclusion. A possible reason could be a difference in 
the lepton reconstruction criteria, such as the isolation parameters, which are not fully 
specified in the CMS study.

\subsubsection{Implementation of the CMS Left-right Symmetric/Lepto\-quark\\ Search at 13\,TeV}
Slepton LSPs produce final states involving both charged leptons and jets, therefore 
existing leptoquark searches might be relevant to gain sensitivity to this class. The search in Ref.\,\cite{CMS:2018iye} looks for final states 
comparable to the $\tilde{\tau}$ ones in Table\,\ref{tab:nv2slep_UDD} and provides 
sufficient information to be implemented in \texttt{CheckMATE\;2}. The search utilizes $\SI{35.9}{\per\femto\barn}$ of proton-proton collision 
data at $\sqrt{s}=\SI{13}{\tera\electronvolt}$. It considers two distinct BSM scenarios.
The first is a left-right symmetric model that introduces new bosons $W^\pm_R, Z^\prime$ and 
heavy neutrinos $N_\ell$, of which $Z^\prime$ and $N_{e,\mu}$ are assumed to be decoupled.
The second model has generic scalar third-generation leptoquarks.
Both models predict final states of the form $\tau\tau jj$, either by single $W_R^\pm$ 
production with the subsequent cascade decay $W_R \rightarrow \tau N_\tau \rightarrow \tau (\tau q \bar{q}^\prime)$, or by pair-produced leptoquarks, which each decay directly into $\tau b$.
We implemented the search into \texttt{CheckMATE\;2} and validated it for the left-right symmetric scenario.
To that end, we used the UFO file of an effective left-right symmetric model provided by Ref.\,\cite{Mattelaer:2016ynf} and available on the \texttt{FeynRules} website.

Unfortunately, we did not find any improvements in the recasted UDD limits compared to vanilla \texttt{CheckMATE\;2}, so we avoid further unnecessary details about the implementation at this point.

\subsection{Framework for Numerical Recasting}
\label{ssec:setup}

We now describe the applications of our framework using benchmark scenarios. 
Similar to our previous work\,\cite{Dreiner_2023}, we assume one non-zero RPV 
coupling at a time for each benchmark. We consider various cases with different 
SUSY particles (sparticles) as LSP, shown in Tables\,\ref{tab:gluino_UDD}, \ref{tab:squark_UDD}, \ref{tab:Bino_UDD}, \ref{tab:W_UDD}, \ref{tab:H_UDD}, and \ref{tab:nv2slep_UDD}. 
Results are presented for two cases: direct pair production of the LSP, and an indirect 
production of the LSP through cascade decays from other sparticles. In each case, the final states differ in the number of final state jets/leptons/missing energy (despite the LSP being the same sparticle) and thus have a distinct signature. All mediating sparticles are assumed to be heavy and decoupled.

\subsubsection{Computational Setup}
We employ the following method to calculate our mass limits: For pair production 
of sparticles, we generate the process at leading order with \texttt{MadGraph5\_aMC@NLO} \cite{Alwall:2014hca} 
using the \texttt{RPVMSSM\_UFO}\,\cite{rpvmssm} model file. 
For the direct production case, the LSPs are pair-produced through $pp$ collisions. 
For cascade decays, we first pair-produce sparticles through direct/associated 
channels, and then allow for the sparticles to decay into the LSP. Until this point,
all processes follow the usual MSSM Feynman rules. The LSPs then decay promptly 
according to the RPV coupling in each considered benchmark.  
For the cascade decays, the two-body decays of sparticles to LSPs are computed by \texttt{MadGraph5\_aMC@NLO}.
The RPV decays and showering are then handled by \texttt{Pythia 8.2}\,\cite{Sjostrand:2014zea}.
Once the final decayed and showered samples are produced, these are passed through \texttt{CheckMATE\;2}. \texttt{DELPHES 3}\,\cite{deFavereau:2013fsa} performs the detector simulation within the \texttt{CheckMATE\;2} framework.
The \texttt{CheckMATE\;2} employed in our study is improved by the implementation of the ATLAS multijet search and the CMS search with two OSSF leptons and jets search, see Secs. \ref{sssec:ATLAS_multijet} and \ref{sssec:CMS_leptons8tev}.
In the following, we will refer to the current publicly available \texttt{CheckMATE\;2} as \texttt{CM2} and to the one having our newly implemented searches as \texttt{CM2-mod}.

\subsubsection{Cross Sections}
\label{sssec:crosssections}
The cross sections obtained from \texttt{MadGraph5\_aMC@NLO} are only at NLO accuracy. 
To obtain state-of-the-art results, we will use NNLL cross sections for more 
accurate results. Since the cross sections are just shifted by a $k$-factor to a 
good approximation, it should not affect the event distribution obtained from 
\texttt{MadGraph5\_aMC@NLO}. Hence, the events can be used as input for 
\texttt{CM2} and \texttt{CM2-mod}. 

Sparticle production cross sections at the NNLL level for the colored sector are obtained with the 
\texttt{NNLL-fast} code\,\cite{Beenakker:2016lwe, Beenakker:2024jwh}.
These cross sections were directly used in the $\tilde{g}\tilde{g}$ pair production with decoupled squarks.
The production cross sections involving light squarks cannot be used directly because they are usually given assuming an 8- or 10-fold mass degeneracy.
In this case all $\tilde{q}_{L/R}$ states, only excluding the bottom and/or top flavor, are in the accessible spectrum with the same mass $m_{\tilde{q}}=m_{\tilde{q}_L}=m_{\tilde{q}_R}$.
In our case we want to study a non-degenerate spectrum with only one non-decoupled squark flavor, although we keep the phenomenologically well motivated left-right degeneracy for the first two generations ($u, d, s, c$), wherever the final state is the same for the left- and right-handed squarks.
To achieve this, we use the prescription proposed in
Ref.\,\cite{Beenakker:2024jwh}: We rescale the degenerate 
\texttt{NNLL-fast} cross section for the production of sparticles 
$k,l = \tilde{q}^{(\ast)}, \tilde{g}, \tilde{t}^{(\ast)}$ (with $^\ast$ indicating the complex conjugated field) according to
\begin{equation}
    \label{eq:rescaling}
    \sigma^{\texttt{NNLL-fast}\text{, non-deg.}}_{pp \rightarrow kl} = R_\text{non-deg.} \times \sigma^{\texttt{NNLL-fast}}_{pp \rightarrow kl}\,,
\end{equation}
where the rescaling factor is given by
\begin{equation}
    \label{eq:R_rescaling}
    R_\text{non-deg.} \equiv \frac{\sigma^{\text{LO,non-deg.}}_{pp \rightarrow kl} (m_{\tilde{u}_L}, m_{\tilde{u}_R}, m_{\tilde{d}_L}, m_{\tilde{d}_R}, ...)}{\sigma^\text{LO, deg.}_{pp \rightarrow kl}(m_{\tilde{q}})}\,.
\end{equation}
The LO cross sections can straightforwardly be obtained from \texttt{MadGraph5\_aMC@NLO}. Eqs. \eqref{eq:rescaling} and \eqref{eq:R_rescaling} essentially represent a rescaling of the non-degenerate LO cross section with a $k$-factor relating the degenerate LO and \texttt{NNLL-fast} cross sections.
In general, the $k$-factors are flavor-dependent. For example, as demonstrated in Refs.\,\cite{Gavin:2013kga, Gavin:2014yga}, the $k$-factors relating LO and NLO cross sections range in flavor- and chirality-dependent subchannels between ${\sim} 1.1$ and ${\sim} 1.6$, while the sum of all subchannels is scaled by a $k$-factor of ${\sim} 1.3$.
By explicitly checking we found that the non-degenerate bounds we obtained are not very sensitive to a variation of the $k$-factor around the degenerate LO-NNLL $k$-factor, if the individual subchannel $k$-factors are of similar size as the LO-NLO ones.
The variations in our bounds were ${\approx} \SI{50}{\giga\electronvolt}$, which is a relative change of a few percent for an $\mathcal{O}(\SI{}{\tera\electronvolt})$ exclusion bound.

For the electroweakino and slepton production cross sections, we compute the cross sections at the NLL level using the \texttt{Resummino}\,\cite{Fiaschi_2023} code.

\section{Results}
\label{sec:results}

\subsection{Direct Production}
\label{ssec:direct}

\begin{figure}[hbt!]
    \centering
    \includegraphics[width=\textwidth]{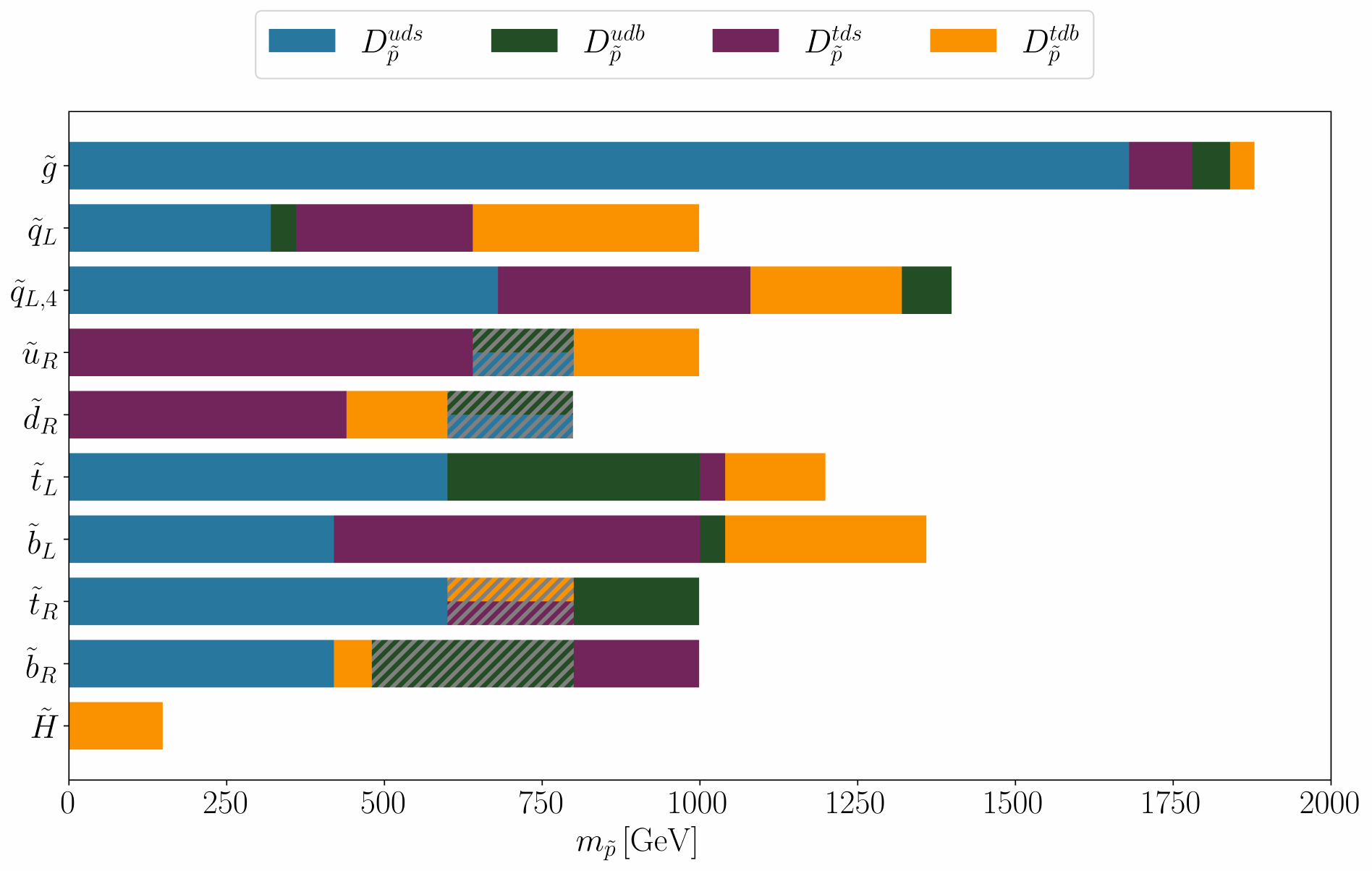}
    \caption{Gluino, squark and slepton LSP direct production search limits at 95\% confidence level with the LSP decaying through UDD couplings. All results are shown for the conservative limit, except the $\tilde{q}_{L,4}$, which stands for $\tilde{q}_{L}$ with 4-fold degeneracy. The $\tilde{p}$ in the legend stands for the relevant LSP on the y-axis. The CMS search result\,\cite{CMS:2022usq} is  shown in hashed-gray. The exact numbers can be found in Table\,\ref{tab:allLSPssearches}, in cases of multiple mass exclusion limits the most sensitive one is shown.}
    \label{fig:direct-lsp}
\end{figure}

In this section, we provide details for the 95\% confidence level
mass exclusion limits for direct production of various LSP 
scenarios. The results are summarized in Fig.\,\ref{fig:direct-lsp}. The final states for each LSP after 
decay through a UDD coupling are provided in Tables\,\ref{tab:gluino_UDD}--\ref{tab:nv2slep_UDD}. As mentioned before, in each benchmark, all sparticles except the LSP
are considered to be decoupled, and the LSP decays promptly. 
The requirement of the presence of $b$-tagged jets results in a higher background rejection.
Therefore, the couplings $\lam''_{113}$ and $\lambda''_{313}$ yield stricter mass exclusion limits in general, as compared to the other couplings.
In the gluino and squark LSP scenario, the best results are obtained employing the $\SI{13}{\tera\electronvolt}$ searches, while currently, for 
the slepton LSP case the $\SI{8}{\tera\electronvolt}$ searches 
provide the best exclusion limits.

Since the pair production cross section for the bino LSP is small, the mass exclusions for these decays are not presented for direct production. Instead, we present results in the next subsection for cascade decays, where the bino LSP is produced through another SUSY particle.\\

\noindent
\textbf{Gluino LSP:} For gluinos, the final states always contain $\geq6$ jets, as can be seen in Table\,\ref{tab:gluino_UDD}. 
Hence, the best results are obtained using the implemented ATLAS multijet search. The high pair production cross section of gluinos at the LHC leads to exclusions close to the kinematic limit, which are the best among all the possible LSPs.
We use the production cross sections obtained from the \texttt{NNLL-fast 2.0} code\,\cite{Beenakker:2024jwh}. Gluino 
masses as high as ${\sim} \SI{1850}{\giga\electronvolt}$ (for $\lambda''_{313}$) can be excluded, as seen in Fig.\,\ref{fig:direct-lsp}.\\

\noindent
\textbf{Squark LSP:} Squark LSP scenarios are divided into the following categories: left-handed first or second generation squarks $\tilde{q}_L$, two right-handed first or second generation squarks $\tilde{u}_R$ and $\tilde{d}_R$, and the four third generation squarks $\tilde{t}_L$, $\tilde{t}_R$, $\tilde{b}_L$, and $\tilde{b}_R$. The possible final states for each scenario are 
shown in Table\,\ref{tab:squark_UDD}. We show limits for each squark\,\footnote{We imply that the results apply to each of the individual squarks in $\tilde{q}_L$, where only one of them is pair-produced.} in Fig.\,\ref{fig:direct-lsp}, assuming a single non-degenerate light squark in the spectrum. 
Additionally, we also show results for the case of a 4-fold degeneracy between the first or second generation squarks $\tilde{q}_L$
($\tilde{u}_L, \tilde{d}_L, \tilde{c}_L$ and $\tilde{s}_L$), and label it as $\tilde{q}_{L,4}$. Each of the above-mentioned scenarios shows varying exclusions due to the presence or absence of $b-$jets and the number of final state jets. 
$\tilde{q}_{L,4}$, naturally, shows the best exclusion limit, since the ATLAS multijet analysis signal regions are well populated by the jets coming from $\tilde{q}_L$ and the 4-fold enhanced production cross section. For the right-handed squarks 
and the third generation squarks, specifically for the 4-jet final state topologies, other studies show better sensitivity. However, the exclusion limits are weaker. Table\,\ref{tab:allLSPssearches} provides the relevant searches and signal regions that are sensitive for each obtained mass exclusion. Note, for 
some of the cases (for example, $\tilde{b}_L$ with $\lambda''_{113}$), there are two possible decay modes of the LSP for a given UDD coupling. In such cases, we
assume a 50\% branching fraction to each of the decay modes for the results in Fig.\,\ref{fig:direct-lsp}. We also show results for 
100\% branching fraction for each decay mode in Appendix\,\ref{app:plots}, Fig.\,\ref{fig:direct_tRbR_branched}.

For completeness, we mention an improved exclusion mass limit for
$\tilde{t}_R$ from a study by CMS\,\cite{CMS:2022usq}. This 
exclusion limit is higher in comparison to the one obtained from \texttt{CM2} for $\lambda_{312}^{''}$. The search is inclusive in its final states, and therefore, the exclusion bound can also be translated for benchmarks with $4j$ final states where $j$ can also be a $b$-jet.
Thus, this limit can be applied to
the $\tilde{u}_R$ and $\tilde{d}_R$ squark LSPs in the $\lam_{112}^{''}$ and $\lam_{113}^{''}$ cases, for $\tilde{b}_R$ in the $\lam_{113}^{''}$ case and $\tilde{t}_R$ in the $\lam_{313}^{''}$ case. These are 
shown in gray-hashed in Fig.\,\ref{fig:direct-lsp}.
One can expect that demanding $b$-tagged jets in final states involving $b$ quarks would give improved mass exclusions.\\

\noindent
\textbf{Slepton LSP:} Slepton LSP pair production leads to 2 leptons
+ 6 jets in the final state, where the jets are of any type. 
However, the lepton(s) can also be neutrinos, leading to missing 
transverse energy. Due to the leptons in the final state, the 
ATLAS multijet search is not sensitive to sleptons. 
Tables\,\ref{tab:nv2slep_UDD} and \ref{tab:nv3slep_UDD} present all 
the possible final states for the cases of a slepton LSP. We use the
\texttt{Resummino}\,\cite{Fiaschi_2023} code to obtain the slepton 
pair production cross section.
It can provide NLL cross sections for slepton pair production at
$\SI{8}{\tera\electronvolt}$ and $\SI{13}{\tera\electronvolt}$ 
center-of-mass energies. With the current version of
\texttt{CM2}, we find no exclusion for sleptons. The search 
\textbf{cms\_exo\_14\_014} is implemented in \texttt{CM2}, but 
requires same sign dileptons, which does not apply to our case. The
CMS search \textbf{cms\_1602\_04334} at 8\,TeV LHC, implemented by 
us in \texttt{CM2-mod}, still provides the best sensitivity for 
selectrons and smuons, but is not enough to exclude any region of 
the parameter space. An update of this or a similar search at the 
13\,TeV LHC might provide bounds for sleptons.

Searches have been performed with similar final states, such as with 1 lepton + 6 jets\,\cite{ATLAS:2024xkd, ATLAS:2023pja, ATLAS:2023cbt, ATLAS:2023bzb, ATLAS:2023mcc} and 2 leptons + 6 jets\,\cite{CMS:2019wav, ATLAS:2022ihe, CMS:2018iye}. 
However, their implementation in \texttt{CheckMATE\;2} is not 
straightforward without additional details, or they apply specific 
cuts for signals like leptoquarks, which are not suitable for this 
particular scenario. For example, we implement the CMS leptoquark 
search\,\cite{CMS:2018iye}, however, we find no improvements in the 
sensitivity for tau sleptons.
Neural-network-based discriminator searches are especially cumbersome to implement in \texttt{CM2} without proper input from experimental collaborations regarding the trained model files. This makes recasting these searches difficult.\\

\begin{table}[hbt!]
    \centering
    \resizebox{\textwidth}{!}{
    \begin{tabular}{c|c|c|c|c}
    \hline
    \hline
        LSP & Coupling  & Sensitive Search & Signal Region & Mass Exclusion (GeV)\\
        &($\lam''_{ijk}$)&&& \\
        
        \hline 
        \hline
     \rule{0pt}{1.0\normalbaselineskip}$\tilde{q}_L $ &112&\textbf{atlas\_1807\_07447}\,\cite{ATLAS:2018zdn} &\textbf{7j}&320 \\
       \rule{0pt}{1.0\normalbaselineskip}$\tilde{q}_L $ &113&\textbf{atlas\_1807\_07447}\,\cite{ATLAS:2018zdn} &\textbf{2b5j}&360 \\
         \rule{0pt}{1.0\normalbaselineskip}$\tilde{q}_L/ \tilde{u}$ &312&\textbf{atlas\_2106\_09609}\,\cite{ATLAS:2021fbt}&\textbf{SR16/SR14}&640 \\
         \rule{0pt}{1.0\normalbaselineskip}$\tilde{q}_L/ \tilde{u}$ &313&\textbf{atlas\_2106\_09609}\,\cite{ATLAS:2021fbt}&\textbf{SR4}&1000 \\
  \rule{0pt}{1.0\normalbaselineskip}$\tilde{q}_{L,4}$ &112&\textbf{atlas\_1807\_07447}\,\cite{ATLAS:2018zdn} &\textbf{9j}&680 \\
   \rule{0pt}{1.0\normalbaselineskip}$\tilde{u}_R / \tilde{d}_R$ &112/113&\textbf{cms\_2206\_09997}\,\cite{CMS:2022usq} &-&800 \\
        \rule{0pt}{1.0\normalbaselineskip}$\tilde{u}_R / \tilde{d}_R$ &113&\textbf{atlas\_1807\_07447}\,\cite{ATLAS:2018zdn} &\textbf{2b2j}&380 \\
        \rule{0pt}{1.0\normalbaselineskip}$\tilde{d}_R$ &312&\textbf{atlas\_1807\_07447}\,\cite{ATLAS:2018zdn}&\textbf{MET1b5j}&440 \\
        \rule{0pt}{1.0\normalbaselineskip}$\tilde{d}_R$ &313& \textbf{atlas\_1807\_07447}\,\cite{ATLAS:2018zdn}&\textbf{4b2j}&600 \\\hline
          \rule{0pt}{1.0\normalbaselineskip}$\tilde{b}_L / \tilde{b}_R$ &112&\textbf{atlas\_1807\_07447}\,\cite{ATLAS:2018zdn} &\textbf{2b6j}&420 \\
        \rule{0pt}{1.0\normalbaselineskip}$\tilde{b}_L$ &113&\textbf{atlas\_2106\_09609}\,\cite{ATLAS:2021fbt}&\textbf{SR4}&1040 \\
             \rule{0pt}{1.0\normalbaselineskip}$\tilde{b}_L / \tilde{b}_R$ &312&\textbf{atlas\_2106\_09609}\,\cite{ATLAS:2021fbt}&\textbf{SR4}&1000 \\
          \rule{0pt}{1.0\normalbaselineskip}$\tilde{b}_L$ &313&\textbf{atlas\_2106\_09609}\,\cite{ATLAS:2021fbt}&\textbf{SR4}&1360 \\  
          \rule{0pt}{1.0\normalbaselineskip}$\tilde{b}_R$ &113&\textbf{atlas\_1807\_07447}\,\cite{ATLAS:2018zdn}&\textbf{4j}&220 \\  
          \rule{0pt}{1.0\normalbaselineskip}$\tilde{b}_R$&113&\textbf{cms\_2206\_09997}\,\cite{CMS:2022usq} &-&800 \\
          \rule{0pt}{1.0\normalbaselineskip}$\tilde{b}_R$ &313&\textbf{atlas\_1706\_03731}\,\cite{ATLAS:2017tmw}&\textbf{Rpv2L1bS}&480 \\
          \rule{0pt}{1.0\normalbaselineskip}$\tilde{t}_L/\tilde{t}_R$ &112&\textbf{atlas\_2004\_10894}\,\cite{ATLAS:2020qlk}&\textbf{Cat2}&600 \\
           \rule{0pt}{1.0\normalbaselineskip}$\tilde{t}_L/\tilde{t}_R$ &113&\textbf{atlas\_2106\_09609}\,\cite{ATLAS:2021fbt}&\textbf{SR10}&1000 \\
           \rule{0pt}{1.0\normalbaselineskip}$\tilde{t}_L$ &312&\textbf{atlas\_2106\_09609}\,\cite{ATLAS:2021fbt}&\textbf{SR4}&1040 \\
           \rule{0pt}{1.0\normalbaselineskip}$\tilde{t}_L$ &313&\textbf{atlas\_2106\_09609}\,\cite{ATLAS:2021fbt}&\textbf{SR10}&1200 \\
           \rule{0pt}{1.0\normalbaselineskip}$\tilde{t}_R$ &312&\textbf{atlas\_1807\_07447}\,\cite{ATLAS:2018zdn}&\textbf{MET5j}&280 \\
           \rule{0pt}{1.0\normalbaselineskip}$\tilde{t}_R$&312/313&\textbf{cms\_2206\_09997}\,\cite{CMS:2022usq} &-&800 \\
           \rule{0pt}{1.0\normalbaselineskip}$\tilde{t}_R$ &313&\textbf{atlas\_1807\_07447}\,\cite{ATLAS:2018zdn}&\textbf{2b2j}&400 \\\hline
           \rule{0pt}{1.0\normalbaselineskip} $\tilde{H}$ &313& \textbf{atlas\_1807\_07447}\,\cite{ATLAS:2018zdn}&\textbf{MET2b1j}&150 \\
         \hline
         \hline
    \end{tabular}
    }
    \caption{List of relevant ATLAS and CMS searches and the corresponding dominant signal regions for the direct production exclusion limits shown in Fig.\,\ref{fig:direct-lsp}.
    For the cases of $\tilde{g}$ LSPs (all four couplings) and $\tilde{q}_{L,4}$ LSPs (for $\lambda''_{113}, \lambda''_{312}, \lambda''_{313}$), the ATLAS multijet, \textbf{atlas\_2401\_16333}, is the most sensitive search.
    The search \textbf{cms\_2206\_09997} is not implemented in \texttt{CheckMATE\;2}, we directly use the result here.
    We obtain no exclusions from \texttt{CM2}/\texttt{CM2-mod} for the rest of the LSPs and benchmark couplings, not shown in this table.}
    \label{tab:allLSPssearches}
\end{table}

\noindent
\textbf{Electroweakino LSP:} For wino LSPs, we assume degenerate 
$\tilde{\chi}_1^0$ and $\tilde{\chi}_1^\pm$ produced
directly at the LHC. For higgsino LSPs, we consider the processes
$pp\to \tilde{\chi}_1^0\tilde{\chi}_2^0$, $pp\to \tilde {\chi}_2^0 
\tilde{\chi}_1^\pm$, and $pp\to \tilde{\chi}_1^+\tilde{\chi}_1^-$ 
for degenerate $\tilde{\chi}_1^0$, $\tilde{\chi}_2^0$ and $\tilde 
{\chi}_1^\pm$. The production cross sections are obtained using 
\texttt{Resummino}\,\cite{Fiaschi_2023} with 
\SI{13}{\tera\electronvolt} center-of-mass energy. For the winos and
higgsinos, we assume equal decay branching fractions to charged 
and neutral current final states, which are shown in 
Tables\,\ref{tab:W_UDD} and \ref{tab:H_UDD}, and consider only the shortest cascades.\,Due to the small 
pair production cross-section of the electroweakinos, along with the presence of vector bosons in some of the final states, the searches implemented in \texttt{CM2-mod} 
cannot provide a mass exclusion\,\footnote{The decay of wino-like
LSPs via UDD operators always involve vector bosons, which 
can decay hadronically or leptonically. However, all the signal 
regions of the ATLAS multijet search require a minimum $p_T$ of 
180\,GeV for the signal jets, which is difficult to get from vector boson decays. The search also vetoes leptons. Moreover, the production cross sections of both the wino and higgsino-like LSPs are very small. This is in addition to the already small signal efficiency due to strong analysis cuts.}. 
We obtain no exclusion for the winos with 
\texttt{CM2}. A weak limit at \SI{150}{\giga\electronvolt} for 
higgsinos is obtained in the $\lam''_{313}$ case with \texttt{CM2}, 
shown in Fig.\,\ref{fig:direct-lsp}. Ref.\,\cite{Barman:2020azo} 
studies the projected sensitivity of electroweakino LSPs decaying 
via UDD couplings at the HL-LHC. A broad overview of CMS searches 
for similar types of final states can be found in 
Ref.\,\cite{CMS:2024bni}.

\subsection{Production from Cascade Decays}
\label{ssec:cascade}

If the LSPs may not be directly produced, we shall consider the possibility that they result from 
the cascade decay of a heavier particle. Here, we focus on cases where a non-LSP sparticle is pair-produced
(or singly-produced along with
the LSP) through $pp$ interactions, and then the sparticle 
decays to the LSP through a cascade. The final states in such a case would be 
distinct from the case of direct LSP 
production, and in some cases an improved exclusion limit can be 
achieved through recasting of different ATLAS/CMS searches. 

We now present benchmark scenarios where cascade 
decays provide better results, \textit{i.e.} stricter mass 
bounds. We denote these 
benchmark scenarios with the label $I_{\tilde
{s}\to\tilde{p}}$, where $\tilde{s}$ and $\tilde{p}$ are the NLSP 
and LSP, respectively. For each cascade benchmark scenario $I_ 
{\tilde{s}\rightarrow\tilde{p}}$, the final states are the same as for 
$D_{\tilde{p}}$, plus extra jets. The number and flavor of the extra
jets depends on the NLSP $\tilde{s}$.

\subsubsection*{Squark LSPs from a Gluino Cascade}

Gluinos have the highest direct pair production cross section. Thus, one can expect to obtain better exclusion limits when an LSP squark is produced through the decay of a gluino NLSP. 
\paragraph{$\boldsymbol{I_{\tilde{g}\rightarrow\tilde{q}_L}}:$}
We first study the case where the $\tilde{g}$ can decay into first- or second-generation squarks.
This can happen in two ways: one can have pair-produced gluinos 
that both decay to the squark LSP, \textit{e.g.} $\tilde{g}\tilde{g}
\to \tilde{q}_L\tilde{q}_L+X$, where $X$ 
indicates further non-supersymmetric particles in the decay,
or associated production where we first directly produce $\tilde{g}\tilde{q_L}$, and then the
gluino further decays to the squark LSP $+X$. The cross
section for the associated production depends heavily on the LSP 
generation due to PDF suppression of the higher generation quarks. The associated production also
dominates the overall cross section. For example, the $\tilde{g} 
\tilde{u}_L$ cross section is at least ${\sim} 5$ times higher in contribution than the direct $\tilde{g}\tilde{g}$ cross 
section, even for only slightly heavier gluinos.
For a large mass difference between the squark and the gluino, the $\tilde{g} 
\tilde{u}_L$ cross section is almost two orders of magnitude higher than the direct $\tilde{g}\tilde{g}$ cross section.
In this work, we consider the best case scenario, where $\tilde{q}=\tilde{u}_L$ only. This is due to the larger $u$-quark PDF in the proton.
%
%
We sum over the cross sections of the direct and associated 
production, details of the cross section calculation are provided
in Section\,\ref{sssec:crosssections}. We scan over a range of gluino and squark masses such that $m_{\tilde{g}}> m_{\tilde{u}_L}$. 

\begin{figure}[hbt!]
    \centering
    \includegraphics[width=0.60\linewidth]{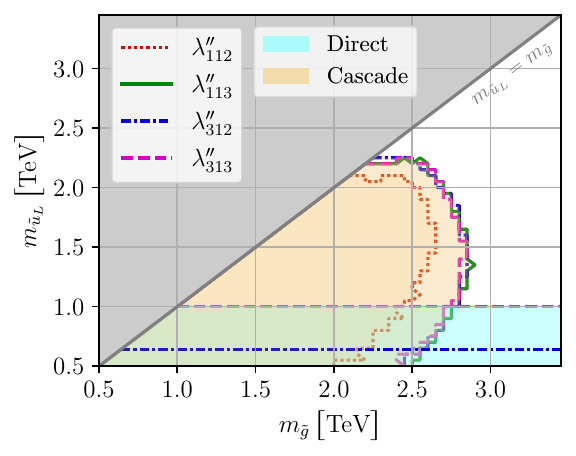}
    \caption{Exclusion regions (in shades of {\it orange}) corresponding to 95\% confidence level for the $\tilde{g}$ to $\tilde{u}_L$ cascade decays $I_{\tilde{g}\rightarrow\tilde{u}_L}$. The bounds of Fig.\,\ref{fig:direct-lsp} from direct squark production $D_{\tilde{q}_L}$ also apply to the scenario and are shown in shades of {\it blue}.
    The {\it gray} region is kinematically disallowed in the scenario. The {\it red} dotted, {\it green} solid, {\it blue} dot-dashed and {\it magenta} dashed contours correspond to couplings $\lambda''_{112}$, $\lambda''_{113}$, $\lambda''_{312}$ and $\lambda''_{313}$, respectively.}
    \label{fig:uL_LSP}
\end{figure}

Fig.\,\ref{fig:uL_LSP} shows the mass exclusion range for this 
scenario $I_{\tilde{g}\rightarrow\tilde{u}_L}$. We also show the results from direct production of $\tilde{q}_L$ for comparison. Due to the presence 
of $t$- and $b$-jets in the final state, the benchmark scenarios corresponding to the couplings $\lam''_{113}, \lam''_{312},$ and $ \lam''_{313}$ have stricter exclusions than the $\lam''_{112}$ case. The ATLAS multijet search shows the maximum sensitivity over the whole range of the mass plane. For the coupling $\lam''_{112}$, the sensitive signal
regions are \textbf{SR1}, \textbf{SR2}, \textbf{SR3}, \textbf{SR4}, and \textbf{SR5}. For all other couplings, the 
signal region \textbf{SR1bj} is most sensitive.

\begin{figure}[hbt!]
    \centering
    \includegraphics[width=0.49\linewidth]{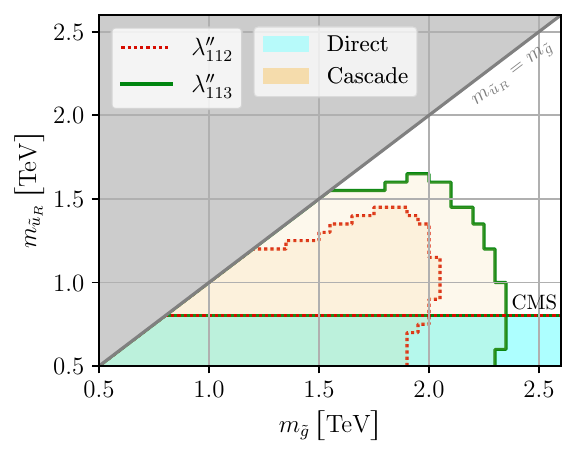}\,
    \includegraphics[width=0.49\linewidth]{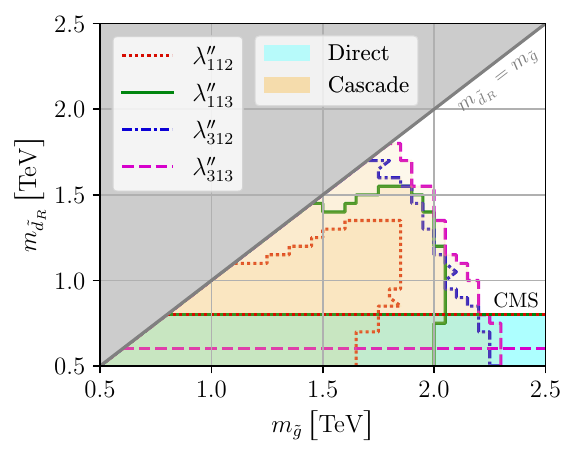}
    \caption{Same as Fig.\,\ref{fig:uL_LSP}, for $I_{\tilde{g}\rightarrow\tilde{u}_R}$ (\textit{left}) and $I_{\tilde{g}\rightarrow\tilde{d}_R}$ (\textit{right}). For the $I_{\tilde{g}\rightarrow\tilde{u}_R}$ cascade, the results for $\lambda''_{312}$ and $\lambda''_{313}$ will be the same as $I_{\tilde{g}\rightarrow\tilde{u}_L}$.}
    \label{fig:uR_dR_LSP}
\end{figure}

\paragraph{$\boldsymbol{I_{\tilde{g}\rightarrow\tilde{u}_R}}$ \textbf{and} $\boldsymbol{I_{\tilde{g}\rightarrow\tilde{d}_R}:}$}
For the right-handed up-type squark LSPs, Table\,\ref{tab:squark_decays} shows that the LSP will decay directly to two light-flavor jets for $\lam''_{112}$, and to a light-flavor jet and a 
$b$-tagged jet for $\lam''_{113}$. For the other two benchmark couplings, $\tilde 
{u}_R$ decays via a cascade, similar to the $\tilde{u}_L$ LSP. This would lead to the
same final states, with the same production cross section, leading to the same mass exclusions as obtained for the $I_{\tilde{g}\rightarrow\tilde{u}_L}$
cascade within uncertainties. Therefore, we only perform scans 
for the $I_{\tilde{g}\rightarrow\tilde{u}_R}$ cascade for the 
couplings of the type $\lam''_{112}$ and $\lam''_{113}$. The {\it
left} panel of Fig.\,\ref{fig:uR_dR_LSP} shows the  
mass exclusion in the $(m_{\tilde{g}},m_{\tilde{u}_R})$ mass plane 
obtained with \texttt{CM2-mod}. The limit from the direct production of the $\tilde{u}_R$ LSP decaying to a pair of 
light-flavor jets from the CMS search\,\cite{CMS:2022usq} is also shown 
for $\lam''_{112}$. For both $\lam''_{112}$ and $\lam''_{113}$ 
couplings, the strongest exclusion comes from \textbf{atlas\_2401\_16333}, dominantly from the \textbf{SR2} and \textbf{SR1bj} signal regions, respectively. Gluino masses up to 1.9\,TeV (2.3\,TeV) are excluded for a $\tilde{u}_R$ LSP of mass 500\,GeV, for the UDD coupling $\lam''_{112}$ ($\lam''_{113}$).

In the case of the $I_{\tilde{g}\rightarrow\tilde{d}_R}$, for all four
benchmark couplings considered here, the LSP can directly decay 
via the UDD coupling. The {\it right} panel of Fig.\,\ref{fig:uR_dR_LSP} shows the exclusion in the 
$(m_{\tilde{g}},m_{\tilde{d}_R})$ mass plane generated with \texttt{CM2-mod}. The limit from the direct production of 
the $\tilde{d}_R$ LSP is also shown for $\lambda''_{313}$. 
For the $\lam''_{312}$ and $\lambda''_{313}$ couplings, searches implemented in \texttt{CM2} increase the sensitivity compared to the ATLAS multijet search (\textbf{SR1bj}), especially in the region of light $\tilde{d}_R$ and heavy $\tilde{g}$, and the region with 
small $m_{\tilde{g}}-m_{\tilde{d}_R}$ difference, where the searches \textbf{atlas\_1909\_08457} (\textbf{Rpv2L}) and \textbf{atlas\_2101\_01629} (\textbf{6J\_btag\_2800}) provide the best exclusions. Gluino masses up to 2.3\,TeV are excluded for a $\tilde{d}_R$ LSP of mass 500\,GeV for both the UDD couplings $\lam''_{112}$ and $\lam''_{113}$.

\begin{figure}[hbt!]
    \centering
    \includegraphics[width=0.49\linewidth]{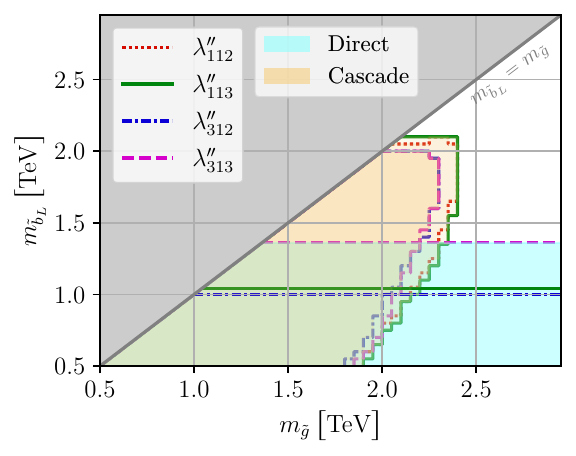}\,
    \includegraphics[width=0.49\linewidth]{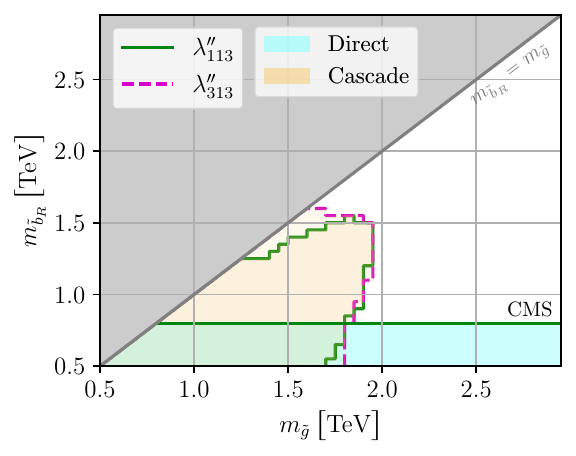}
    \caption{Same as Fig.\,\ref{fig:uL_LSP}, for $I_{\tilde{g}\rightarrow\tilde{b}_L}$ ({\it left}) and $I_{\tilde{g}\rightarrow\tilde{b}_R}$ ({\it right}) decays. For the $I_{\tilde{g}\rightarrow\tilde{b}_R}$ cascade, the results for $\lambda''_{112}$ and $\lambda''_{312}$ will be the same as $I_{\tilde{g}\rightarrow\tilde{b}_L}$.}
    \label{fig:bLR_LSP}
\end{figure}

\paragraph{$\boldsymbol{ I_{\tilde{g}\rightarrow\tilde{b}_L}}$ and $\boldsymbol{I_{\tilde{g}\rightarrow\tilde{b}_R}}$:} 
When $\tilde{b}_L$ or $\tilde{b}_R$ is the LSP, we do not have the associated production channel due to PDF suppression. Therefore, the production cross section is just that of gluino pair production at the LHC. For all the benchmark 
couplings, the $\tilde{b}_L$ LSP has a cascade decay (see Table\,\ref{tab:squark_decays}). In the \textit{left} panel of Fig.\,\ref{fig:bLR_LSP}, the exclusion limits in the 
$(m_{\tilde g},m_{\tilde b_L})$
mass plane for $I_{\tilde{g}\rightarrow\tilde{b}_L}$ are shown. The direct limits for $\tilde{b}_L$ are also shown as horizontal lines. For 
the couplings $\lam''_{112}$, $\lam''_{113}$, and $\lam''_{312}$, \textbf{atlas\_2401\_16333} is the most sensitive search 
throughout the mass plane. For the coupling $\lambda''_{312}$, most of the exclusion also comes from \textbf{atlas\_2401\_16333} (\textbf{SR1bj} and \textbf{SR2bj}), but a few points in the low $\tilde{b}_L$ and high $\tilde{g}$ mass range are excluded by \textbf{atlas\_2106\_09609} (\textbf{SR10}).

The \textit{right} panel of Fig.\,\ref{fig:bLR_LSP} shows 
the mass exclusion results for $I_{\tilde{g}\rightarrow\tilde{b}_R}$. For $\lam''_{113}$, \textbf{atlas\_2401\_16333} (\textbf{SR1bj}) is the most sensitive search throughout. For $\lam''_{313}$, \textbf{atlas\_2401\_16333} shows the highest sensitivity for most of the mass plane, however, a few mass points are also excluded by the searches \textbf{atlas\_1909\_08457} (\textbf{Rpv2L}) and \textbf{atlas\_2109\_01629} (\textbf{6J\_btag\_2800}). The results for $\lam''_{112}$ and $\lam''_{312}$ are the same as in the $(m_{\tilde{g}},m_{\tilde{b}_L})$ plane, since the final states are the same (see Table\,\ref{tab:squark_decays}).

\begin{figure}[hbt!]
    \centering
    \includegraphics[width=0.49\linewidth]{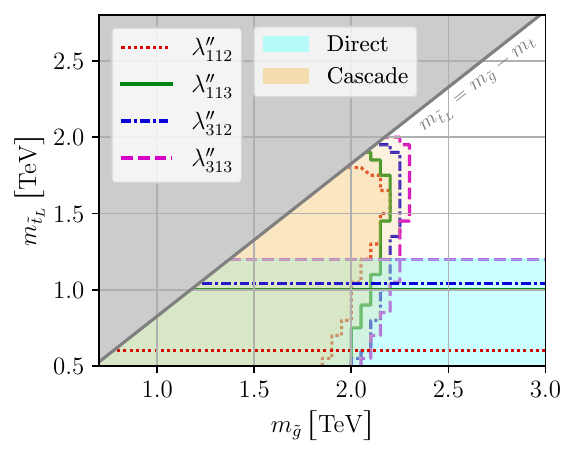}\,
    \includegraphics[width=0.49\linewidth]{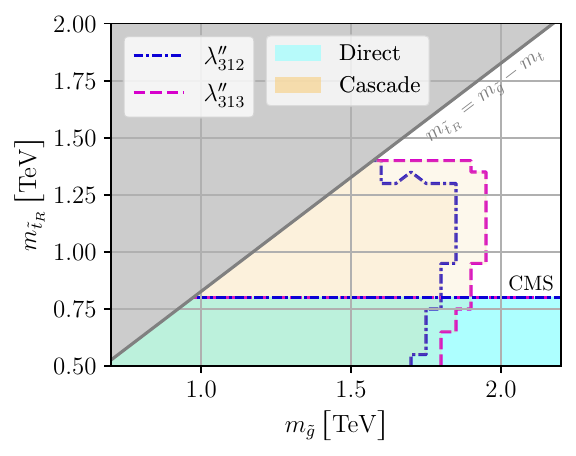}
    \caption{Same as Fig.\,\ref{fig:uL_LSP}, for $I_{\tilde{g}\rightarrow\tilde{t}_L}$ ({\it left}) and $I_{\tilde{g}\rightarrow\tilde{t}_R}$ ({\it right}) decays. For the $I_{\tilde{g}\rightarrow\tilde{t}_R}$ cascade, the results for $\lambda''_{112}$ and $\lambda''_{113}$ will be the same as $I_{\tilde{g}\rightarrow\tilde{t}_L}$.}
    \label{fig:tLR_LSP}
\end{figure}

\paragraph{$\boldsymbol{I_{\tilde{g}\rightarrow\tilde{t}_L}}$ and $\boldsymbol{I_{\tilde{g}\rightarrow\tilde{t}_R}}$:}
Similar to the $I_{\tilde{g}\rightarrow\tilde{b}_L}$ or $I_{\tilde{g}\rightarrow\tilde{b}_R}$ case, the associated production channel for  $I_{\tilde{g}\rightarrow\tilde{t}_L}$ and $I_{\tilde{g}\rightarrow\tilde{t}_R}$ is suppressed. Thus gluino pair production
gives the dominant cross section at the LHC. For all the 
benchmark couplings, the $\tilde{t}_L$ LSP has a cascade decay 
(see Table\,\ref{tab:squark_decays}). 
The {\it left} panel of Fig.\,\ref{fig:tLR_LSP} shows the exclusion in the $(m_{\tilde{g}},m_{\tilde{t}_L})$ mass
plane for $I_{\tilde{g}\rightarrow\tilde{t}_L}$ obtained with \texttt{CM2-mod}. The limits from the direct 
production of the $\tilde{t}_L$ LSP are also shown for each
of the couplings. For all the benchmark couplings, the relevant 
searches are \textbf{atlas\_2401\_16333} (\textbf{SR1bj} and 
\textbf{SR2bj}) and \textbf{atlas\_2106\_09609} (\textbf{SR4} and
\textbf{SR6}). Gluino masses up to 1.85\,TeV, 2\,TeV and 2.05\,TeV are excluded in the presence of the couplings $\lam''_{112}$, $\lam''_{113}$ ($\lambda''_{312}$), and $\lam''_{313}$, respectively, for a 500\,GeV $\tilde{t}_L$ LSP.

The $\tilde{t}_R$ LSP decays directly for $\lam''_{312}$ 
and $\lam''_{313}$ couplings. The {\it right} 
panel of Fig.\,\ref{fig:tLR_LSP} shows the exclusion in 
the $(m_{\tilde{g}},m_{\tilde{t}_R})$ mass plane for $I_{\tilde{g}\rightarrow\tilde{t}_R}$ obtained 
with \texttt{CM2-mod}. The limit from the direct production of the 
$\tilde{t}_R$ LSP decaying to a pair of light-flavor jets from the CMS search\,\cite{CMS:2022usq} is also shown for $\lambda''_{312}$. In this scenario, the dominant searches are again \textbf{atlas\_2401\_16333} (\textbf{SR1bj}), \textbf{atlas\_1909\_08457} (\textbf{Rpv2L}), and  \textbf{atlas\_2101\_01629} (\textbf{6J\_btag\_2800}). Additionally, the search \textbf{atlas\_2106\_09609} has sensitivity for a large region of this parameter space with the signal region \textbf{SR9} (\textbf{SR10} and \textbf{SR4}) for the $\lambda''_{312}$ ($\lambda''_{313}$) couplings. For the LSP mass of 500\,GeV, gluino masses up to 1.7\,TeV and 1.8\,TeV are excluded for the $\lambda''_{312}$ and $\lambda''_{313}$ couplings, respectively.

\subsubsection*{Bino LSP from a Gluino Cascade}

\begin{figure}[hbt!]
    \centering
    \includegraphics[width=0.49\linewidth]{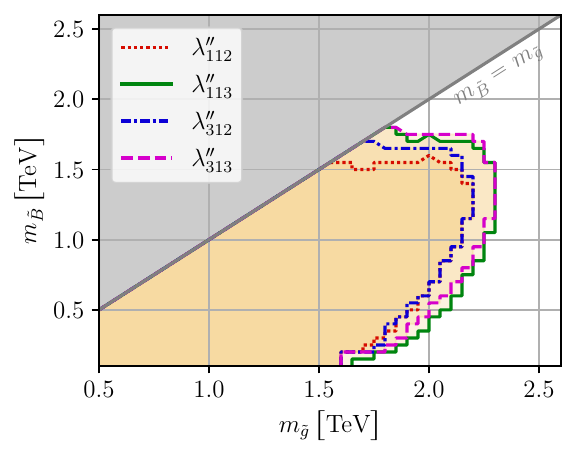}
    \includegraphics[width=0.49\linewidth]{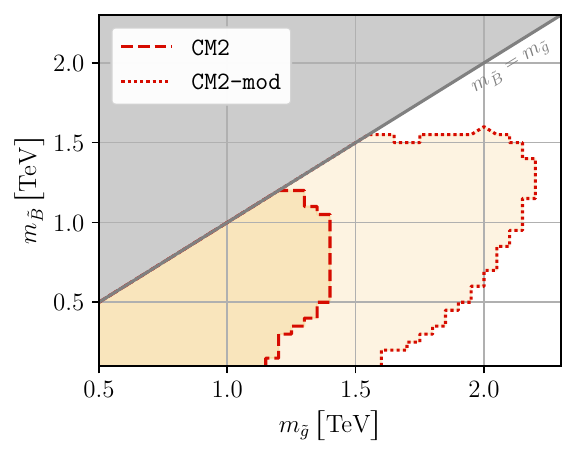}
    \caption{{\it Left:} Same as Fig.\,\ref{fig:uL_LSP}, for  $I_{\tilde{g}\rightarrow\tilde{B}}$; {\it Right:} Comparison of the limits from the already available searches in \texttt{CM2} ({\it red} dashed) and the ATLAS multijet search implemented in \texttt{CM2-mod} ({\it red} dotted) for the $\lambda''_{112}$ coupling for $I_{\tilde{g}\rightarrow\tilde{B}}$.}
    \label{fig:binoLSP_1}
\end{figure}

The direct pair production cross section of bino-like neutralinos is very small, therefore, we only consider its production from the decay of other SUSY particles. We first study the case where the bino is the LSP and gluinos are the NLSPs,  $I_{\tilde{g}\rightarrow\tilde{B}}$. We consider the pair production of gluinos, eventually decaying into the bino-like neutralino LSP, which finally decays to quarks via the various UDD couplings. We scan over a range of gluino and bino masses such that $m_{\tilde{g}}>m_{\tilde{B}}$. The {\it left} panel of Fig.\,\ref{fig:binoLSP_1} shows the exclusion in the gluino-bino mass plane for $I_{\tilde{g}\rightarrow\tilde{B}}$ obtained from \texttt{CM2-mod}.
We find that the edge of the exclusion contour is given by \textbf{atlas\_2401\_16333} for all four UDD couplings. For
the couplings $\lam''_{113}$, $\lam''_{312}$, and $\lam''_{313}$, 
the sensitive signal regions are \textbf{SR1bj} and \textbf{SR2bj}, since the final states include $b$ jets. For $\lam''_{112}$, \textbf{SR2} and \textbf{SR5} play a major role for higher gluino and bino mass.  
However, for the couplings $\lam''_{312}$ and $\lam''_{313}$, the final state top quark is off-shell in the region where the bino mass is below the top quark mass. In that region, the most sensitive search is \textbf{atlas\_2101\_01629}, with the signal region \textbf{6J-btag-2800}.  
For a 100\,GeV bino LSP, the NLSP gluinos are excluded up to a mass of around 1.6--1.65\,TeV, while for a heavier bino of mass of 1.4\,TeV, gluino masses are excluded up to 2.2--2.3\,TeV.
For $\lam''_{112}$ and $\lam''_{113}$, the second most sensitive search from \texttt{CM2} is \textbf{atlas\_1807\_07447}, while 
for $\lam''_{312}$ and $\lam''_{313}$, the major exclusion after 
the ATLAS multijet search comes from \textbf{atlas\_2106\_09609}, with the \textbf{atlas\_1909\_08457} search having some 
sensitivity in both the light $\tilde{B}$ and heavy $\tilde{g}$ region, as well as the region with small $m_{\tilde{g}} - m_{\tilde{B}}$ mass difference. 
The {\it right} panel of Fig.\,\ref{fig:binoLSP_1} shows the extent of improvement achieved for $\lam''_{112}$ by the ATLAS multijet search implemented in \texttt{CM2-mod} for $I_{\tilde{g}\rightarrow\tilde{B}}$, compared to the bare \texttt{CM2} result. We find that \texttt{CM2-mod} improves the sensitivity significantly for $I_{\tilde{g}\rightarrow\tilde{B}}$.

\subsubsection*{Bino LSP from a Squark Cascade}
Next we study the case where the bino LSP is produced from the
cascade decay of squarks. Since the gluinos are decoupled, the 
dominant mechanism of squark pair production is flavor independent.
%
\paragraph{$\boldsymbol{I_{\tilde{q}_{L/R}\to\tilde{B}}\!\!:}$}
For the pair production of one left-right degenerate squark flavor from the first or second generation, we use the 10-fold degenerate \texttt{NNLL-fast} cross sections scaled down by a factor of five.
This is possible since the dominant production process is flavor independent.
We run our analysis with up-squarks $\tilde{u}_{L/R}$, but the same limits apply to $\tilde{d}_{L/R}$, $\tilde{s}_{L/R}$ and $\tilde{c}_{L/R}$.

\begin{figure}[hbt!]
    \centering
    \includegraphics[width=0.60\linewidth]{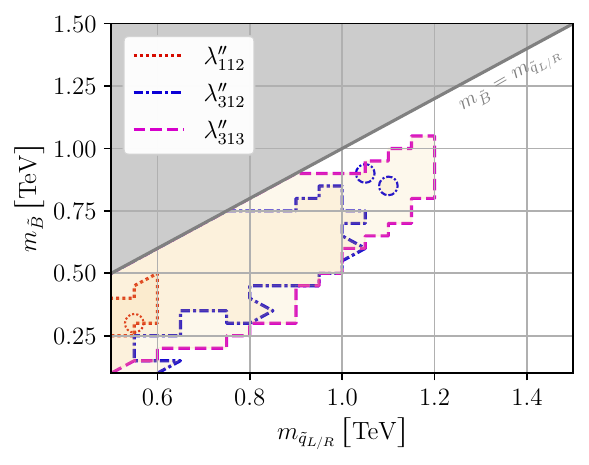}
    \caption{Same as Fig.\,\ref{fig:uL_LSP}, for $I_{\tilde{q}_{L/R}\rightarrow\tilde{B}}$. We use circles in the respective linestyles and colors to indicate fluctuations in the obtained exclusion contours. For $\lambda^{''}_{112}$, the encircled mass point on the exclusion contour is in fact allowed. For $\lambda^{''}_{312}$ there are two encircled points, marking excluded points outside the contour. The results for $\lambda^{''}_{113}$ can be found in Appendix\,\ref{app:plots}, Fig.\,\ref{fig:bino_lsp_st_113} (\textit{left}).}
    \label{fig:binoLSP_q}
\end{figure}

Fig.\,\ref{fig:binoLSP_q} shows the exclusion for $I_{\tilde{q} 
_{L/R}\to\tilde{B}}$ in the $(m_{\tilde{q} _{L/R}},m_{\tilde{B}})$ 
plane, again using \texttt{CM2-mod}. For $\lam^{\prime\prime}_{112}$
the excluded region is small, and
covered by the search \textbf{atlas\_1807\_07447}, with only a 
few points excluded by the multijet search. The excluded 
parameter space for $\lam^{\prime\prime}_{312}$ and $\lam^{\prime
\prime}_{313}$ is much bigger. By far the most sensitive search is \textbf{atlas\_2106\_09609}; only for 
small squark masses, ${\sim} \SI{500}{\giga\electronvolt}$, 
\textbf{atlas\_2004\_10894} yields larger $r$-values for some 
points. In the $\lam^{\prime\prime}_{313}$ case, the 
\textbf{atlas\_2401\_16333} search also yields exclusions, but this 
region is embedded within the \texttt{CM2} exclusion and comes
with lower $r$-values.
For $\lam^{\prime\prime}_{112}$, we encounter one allowed point in the exclusion region that lies in the gap of the two analyses \textbf{atlas\_2401\_16333} and \textbf{atlas\_1807\_07447}. This spoils the marking of a 
proper contour, so we denote it separately with a circle.
For $\lam^{\prime\prime}_{312}$ there are two excluded points 
lying outside the contour, which are caused by the performance of signal regions of \textbf{atlas\_2106\_09609}: Outside the contour (set by \textbf{SR20}), \textbf{SR9} is the most sensitive signal region (with $r < 1$), but for two points, \textbf{SR20} again yields exclusions ($r > 1$).
The excluded points outside the contour are marked with extra circles.
These exceptions usually occur at the transition regions between two different analyses or two different signal regions from the same analysis.

The case $\lam^{\prime\prime}_{113}$ of $I_{\tilde{q}_{L/R}\rightarrow\tilde{B}}$ is shown independently in
Appendix\,\ref{app:plots}, Fig.\,\ref{fig:bino_lsp_st_113} (\textit{left}) as a scatter
plot, because we could not find a clear contour in the lower squark mass region.
The search \textbf{atlas\_2401\_16333} is the most sensitive, but for squark masses 
below ${\sim}\SI{0.8}{\tera\electronvolt}$, the $r$-value obtained 
in the most efficient signal region, \textbf{SR1bj}, jumps for 
adjacent points in parameter space between values bigger and 
smaller than one. 

\paragraph{$\boldsymbol{I_{\tilde{b}_{1}\rightarrow\tilde{B}}}$
and $\boldsymbol{I_{\tilde{t}_{1}\rightarrow\tilde{B}}:}$} In
the case of third-generation squarks, the left-right degeneracy 
is no longer a good approximation because of large off-diagonal terms
in the mixing matrix \cite{Dreiner:2023yus}. Thus, we consider a 
spectrum with a bino LSP and only the light sbottom $\tilde{b}_1$
or stop $\tilde{t}_1$ in the spectrum, while the heavy state 
($\tilde{b}_2$ or $\tilde{t}_2$) is assumed to be decoupled. The
sbottom/stop pair production cross sections were again obtained 
from appropriate rescaling of the \texttt{NNLL-fast} cross 
sections.

\begin{figure}[hbt!]
    \centering
    \includegraphics[width=0.49\linewidth]{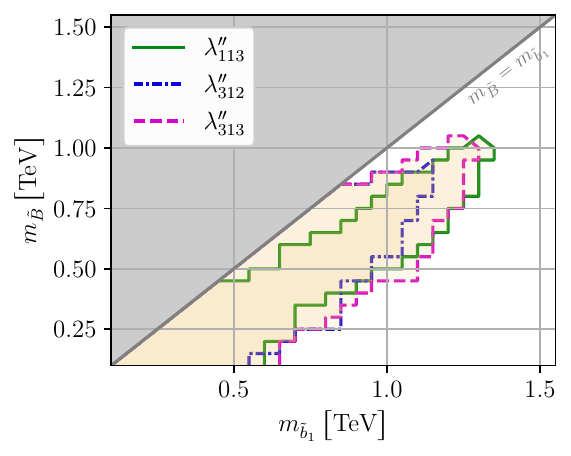}
    \includegraphics[width=0.49\linewidth]{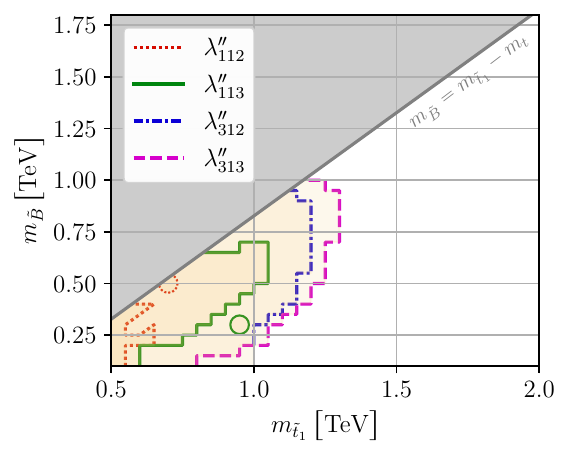}
    \caption{{\it Left:} Same as Fig.\,\ref{fig:uL_LSP}, for $I_{\tilde{b}_{1}\rightarrow\tilde{B}}$; {\it Right:} Same as Fig.\,\ref{fig:uL_LSP}, for $I_{\tilde{t}_{1}\rightarrow\tilde{B}}$. We use circles in the respective linestyles and colors to indicate irregular behaviour of the obtained exclusion contours. For $\lambda^{''}_{112}$ and $\lambda^{''}_{113}$, the encircled points mark exclusions outside the contour. The results for $\lambda^{''}_{112}$ can be found in Appendix\,\ref{app:plots}, Fig.\,\ref{fig:bino_lsp_st_113} (\textit{right}).}
    \label{fig:binoLSP_stsb}
\end{figure}

The \textit{left} panel of Fig.\,\ref{fig:binoLSP_stsb} shows the 
exclusions obtained in the $I_{\tilde{b}_{1}\rightarrow\tilde{B}}$ cascade scenario. For the $\lam^{\prime\prime}_{113}$ case, 
\textbf{atlas\_1807\_07447} excludes the largest region for small sbottom masses, 
while for large masses, \textbf{atlas\_2401\_16333} in 
\textbf{SR1bj} is the most sensitive.
For $\lam^{\prime\prime}_{312}$, \textbf{atlas\_1807\_07447} almost solely yields the exclusion
region, while for $\lam^{\prime\prime}_{313}$, this happens
through \textbf{atlas\_1807\_07447} only for lower masses, and
through \textbf{atlas\_2106\_09609} for the rest of the parameter
space. The results for the $\lambda^{\prime\prime}_{112}$ case 
are not reliable as they do not form a clear exclusion contour, 
however, we still present the results obtained with \texttt{CM2-mod} 
for completeness in the form of a scatter plot in 
Appendix\,\ref{app:plots}, Fig.\,\ref{fig:bino_lsp_st_113} (\textit{right} panel).

The \textit{right} panel of Fig.\,\ref{fig:binoLSP_stsb} shows the exclusions 
obtained in the $I_{\tilde{t}_{1}\rightarrow\tilde{B}}$ cascade scenario. The 
dominant exclusions are due to the searches already implemented in \texttt{CM2}.
For $\lam^{\prime\prime}_{112}$, the small exclusion region in parameter space is 
obtained through \textbf{atlas\_2106\_09609} and \textbf{atlas\_2004\_10894}\,\footnote{Note that the \textbf{atlas\_2004\_10894} search provides sensitivity here even though it requires two photons to be present. We have checked that our signal process passes this 
requirement, even though the process does not involve photons explicitly.} in \texttt{CM2}.
The search \textbf{atlas\_2106\_09609} completely dominates
the exclusion regions for the remaining couplings: $\lam^ 
{\prime\prime}_{113}$, $\lam^{\prime\prime}_{312}$, and 
$\lam^{\prime\prime}_{313}$. We again encounter a few 
points which are excluded and lie outside the large contour; they are 
marked as independent circles.
For $\lambda^{''}_{112}$, the signal region \textbf{SR9} from the search \textbf{atlas\_2106\_09609} is again sensitive outside the contour set by \textbf{SR3}, \textbf{SR9} (\textbf{atlas\_2106\_09609}), and  \textbf{Cat2} (\textbf{atlas\_2004\_10894}).
For $\lam^{''}_{113}$, \textbf{SR10} becomes sensitive outside the contour set by \textbf{SR4} and \textbf{SR10}.
A search by CMS\,\cite{CMS:2021knz,CMS-PAS-SUS-23-001} for 
a $I_{\tilde{t} \rightarrow \tilde{B}}$ type of cascade 
places a lower bound on $\tilde{t}$ mass of \SI{700}
{\giga\electronvolt} for the $\lam''_{112}$ case. This is better than the result obtained using \texttt{CM2}.



\section{Discussions and Conclusion}
\label{sec:concl}

In this work, we extend the framework introduced in 
Ref.\,\cite{Dreiner_2023} to investigate the LLE operators 
and perform a detailed numerical study of the current 
status of UDD operators in the RPV-MSSM. We first 
divide the nine UDD couplings into four sets having similar
final states and select a benchmark coupling from each set.
Next, we identify the final states arising for all possible
LSPs in RPV-SUSY via the 
presence of the four benchmark UDD couplings. The LSPs 
themselves can be directly pair produced at the LHC, or 
they can result from the 
cascade decay via gauge couplings of some other heavier 
sparticles pair produced at the LHC. Our goal is to identify 
potential gaps in the coverage of the UDD couplings. These can
be either due to a relevant experimental search being present, 
but not properly recast in order to apply it to the various RPV scenarios, or very weak sensitivity to specific final states, which must be 
targeted in the future.

We find that very few of the searches relevant for the final 
states in the UDD are implemented in the recasting framework of 
\texttt{CheckMATE\;2}. In order to improve the coverage, we implement three searches in 
\texttt{CheckMATE\;2} for the present study, the ATLAS 13\,TeV  multijet search\,\cite{ATLAS:2024kqk}, the CMS 8\,TeV search for a opposite sign same flavor lepton pair along with jets\,\cite{CMS:2016ooq}, and the CMS 13\,TeV leptoquark search\,\cite{CMS:2018iye}.
Finally, we test the UDD signals with the modified \texttt{CheckMATE\;2} version including these three searches, called \texttt{CM2-mod}.

\begin{table}[hbt!]
    \centering
    \resizebox{\textwidth}{!}{
    \centering
    \begin{tabular}{c | p{2cm} | p{4cm} | c }
    \hline\hline
    \multirow{2}{*}{Search} & Energy, & \multirow{2}{*}{Final state} & \multirow{2}{*}{Signal regions} \\
    & Luminosity &  & \\
    \hline
    \textbf{atlas\_1706\_03731} & 13\,TeV, 36.1\,fb$^{-1}$ & SS $ll$/$\geq 3l$, $\geq 3$-$6j$, $0b$ to $\geq 2b$, $E_T^{\rm miss}$, $m_{\rm eff}$ & \textbf{Rpv2L1bS}\\
    \hline
    \textbf{atlas\_1807\_07447} & \multirow{2}{*}{13\,TeV,} & Model-independent multiple SRs with & \multirow{2}{*}{\textbf{4j, 7j, 9j, 2b2j, 2b5j, 2b6j, 4b2j,}}\\
    & 3.2\,fb$^{-1}$ & $e, \mu, \gamma, j, b, E_T^{\rm miss}$ & \textbf{MET5j, MET1b5j, MET2b1j} \\
    \hline
    \textbf{atlas\_1909\_08457} & 13\,TeV, 139\,fb$^{-1}$ & SS $ll$/$\geq 3l$, $\geq 6j$, $0b$ to $\geq 2j$, $E_T^{\rm miss}$, $m_{\rm eff}$ & \textbf{Rpv2L}\\
    \hline
    \textbf{atlas\_2004\_10894} & 13\,TeV, 139\,fb$^{-1}$ & $2\gamma$(Higgs), $0l$/$\geq 1l$, $\geq 2j$/$< 2j$, $E_T^{\rm miss}$ & \textbf{Cat2}\\
    \hline
    \textbf{atlas\_2101\_01629} & 13\,TeV, 139\,fb$^{-1}$ & $1l$, $\geq 2$-$6j$, $0b$/$\geq 1b$, $E_T^{\rm miss}$, $m_{\rm eff}$ & \textbf{6J\_btag\_2800}\\
    \hline
    \textbf{atlas\_2106\_09609} & 13\,TeV, 139\,fb$^{-1}$ & $1l$/SS $ll$, $\geq 4$-$15j$, $0b$ to $\geq 4b$ & \textbf{SR4, SR9, SR10, SR14, SR16}\\
    \hline
    \textbf{atlas\_2401\_16333} & 13\,TeV, 140\,fb$^{-1}$ & $\geq 7$-$8j$, 0$e/\mu$, 0-2$b$ & All SRs\\
    \hline
    \textbf{cms\_1602\_04334} & 8\,TeV, 19.7\,fb$^{-1}$ & OS $ee/\mu\mu$, $\geq 5j$, $\geq 1b$ & All SRs\\
    \hline
    \textbf{cms\_2206\_09997} & 13\,TeV, 138\,fb$^{-1}$ & $\geq 4j$ & Pairs of dijet resonances \\
    \hline\hline
    \end{tabular}
    }
    \caption{Summary of the sensitive searches found in this study for all possible LSPs decaying via the UDD operators for the LSP produced either directly or from the gauge-cascades of other sparticles.}
    \label{tab:summary_searches}
\end{table}

In Table\,\ref{tab:summary_searches} we present a summary of 
all the relevant searches, along with the final state and 
signal regions, which provide sensitivity for various 
benchmark scenarios studied in this paper. The resulting 
mass bounds from the direct LSP decays are summarized in 
Fig.\,\ref{fig:direct-lsp}. See also
Fig.\,\ref{fig:direct_tRbR_branched} in the Appendix for mass
bounds in 
cases where we have more than one decay mode of the squark 
LSPs. 

RPV SUSY is often considered an attractive avenue for alleviating the naturalness problem due to the weaker limits on the sparticle masses.
However, for that one requires the higgsinos, stop squarks, one sbottom squark and the gluino to be relatively light\,\cite{Bhattacherjee:2013gr,Feng:2013pwa,Buckley:2016kvr}.
With the present LHC searches, although the bounds on higgsinos are still less than 500\,GeV, we find that the gluinos and stops are already excluded up to 1.5-2\,TeV (see Fig.\,\ref{fig:tLR_LSP}). 
It is, therefore, difficult to accommodate naturalness within the present framework.

The results of Fig.\,\ref{fig:direct-lsp} improve
for squark LSPs when we consider their indirect production 
from gluino NLSPs, shown in 
Figs.\,\ref{fig:uL_LSP}--\ref{fig:tLR_LSP}. For bino LSPs, we 
only consider indirect production through gluino and squark 
NLSPs and the results are shown in 
Figs.\,\ref{fig:binoLSP_1}--\ref{fig:binoLSP_stsb}. We 
enumerate our key findings below:
\begin{itemize}
    \item We observe that the UDD colored sector is well covered. 
    Gluino LSPs are excluded up to masses of 1.6--1.85\,TeV, with the most sensitive search being the ATLAS multijet search that we implement in \texttt{CM2-mod}.
    
    \item For squark LSPs, while few LSPs decaying via specific UDD operators are excluded up to masses close to a TeV, some of the scenarios still only have mass bounds below 
    500\,GeV. Experimentalists can particularly target LSP $\to j_l + b$ and $j_l + t$ final states to improve the bounds for some right-handed squark LSPs.
    Even for multijet final states of squark LSPs, the coverage of the ATLAS multijet search is limited for processes having low cross-sections due to the very high $p_T$ requirements on the jets. 
    
    \item We find a gap in the coverage of UDD sleptons, winos, and higgsinos. 
    For sleptons, we observe that the CMS 8\,TeV OSSF lepton pair + jets search still provides the strongest sensitivity, however, is not able to exclude any region of the parameter space.
    An update of this search at the 13\,TeV LHC might yield potential sensitivity.
    The final states in these scenarios are of the kind: 2 OSSF leptons + jets and 2 $V(W/Z/H)$ + jets.
\end{itemize}

The results imply an upgrade of the recasting framework with 
more relevant searches included is necessary to improve the 
coverage of the UDD couplings in the RPV-MSSM.
This is in contrast to the LLE couplings, which have 
comprehensive coverage over all possible LSPs and couplings, 
as deduced in Ref.\,\cite{Dreiner_2023}. New
experimental searches targeting these specific UDD 
generated final states are required. We want to reiterate 
the importance of joint efforts from theorists and 
experimentalists in order to probe the UDD operators of 
RPV-SUSY.

\section*{Acknowledgements}

The authors would like to thank Dominik K\"ohler for help in the initial stages of the project, Saurabh 
Nangia for useful discussions on the ABC of RPV framework, and
Javier Montejo Berlingen for pointing out some of the relevant
ATLAS searches for this study. HKD would like to thank the 
Nikhef Theory Group for their kind hospitality while part of 
this work was completed. NS is supported by the U.S. Department of Energy, Office of Science, Office of High Energy Physics under Award Number DE-SC0011845. 

\clearpage

\appendix


\section{Supplementary Tables and Figures}
\label{app:plots}
While discussing the slepton LSPs, we considered only the two step cascade decay mediated by the bino.
Table\,\ref{tab:nv3slep_UDD} shows the slepton LSP decays for three step cascades.
We don't use them in this work, however we state them here, because they become relevant for non-decoupled higgsinos.

\begin{table}[hbt!]
    \centering
    \resizebox{0.75\textwidth}{!}{
    \centering
    \begin{tabular}{c | c c c}
    \hline\hline
    LSP  &  Coupling  &   LSP Decay  &  Benchmark Label\\
    \hline\hline
        \multirow{3}{*}{$\tilde{e}_L/\tilde{\mu}_L/\tilde{\tau}_L/$}  &    \multirow{2}{*}{$\lambda_{113}''$, $\lambda_{312}''$}  & \textcolor{brown}{$2j_l + 1t + 1V +$ ($e/\mu/\tau/$MET)} & \multirow{2}{*}{$D_{\tilde{e}}^{udb}$, $D_{\tilde{e}}^{tds}$}\\  
        & & \textcolor{brown}{$2j_l + 1b + 1V +$ ($e/\mu/\tau/$MET)} & \\\cline{2-4}
        \multirow{2}{*}{$\tilde{\nu}_e/\tilde{\nu}_\mu/\tilde{\nu}_\tau$}&  \multirow{3}{*}{$\lambda_{313}''$}  & \textcolor{brown}{$1j_l+1b+1t+1V+$ ($e/\mu/\tau/$MET)} &  \multirow{3}{*}{$D_{\tilde{e}}^{tdb}$} \\
        & & \textcolor{brown}{$1j_l+2b+1V+$ ($e/\mu/\tau/$MET)} & \\
        & & \textcolor{brown}{$1j_l+2t+1V+$ ($e/\mu/\tau/$MET)} & \\
    \hline\hline                                                       
    \end{tabular}
    }
    \caption{Details of the left-handed slepton LSP benchmarks when the decay involves a three step cascade (shown in {\it \textcolor{brown}{brown}}), with columns as in Table\,\,\ref{tab:gluino_UDD}.} 
    \label{tab:nv3slep_UDD}
\end{table}

\begin{figure}[hbt!]
    \centering
    \includegraphics[width=0.49\linewidth]{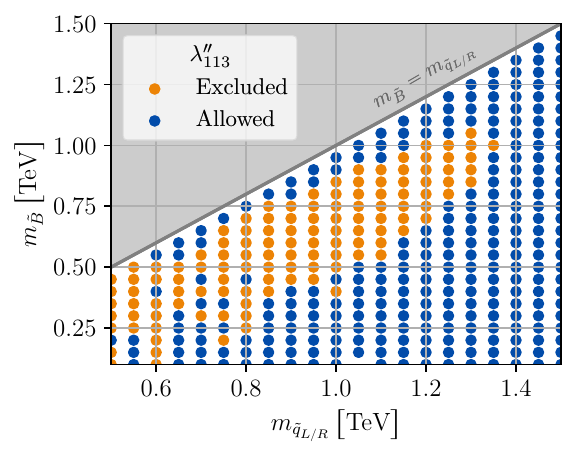}
    \includegraphics[width=0.48\linewidth]{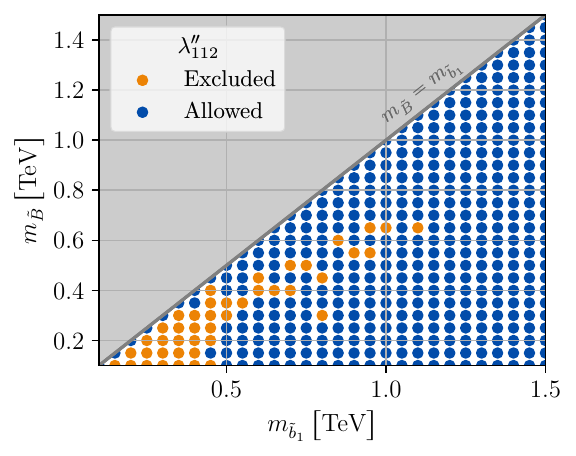}
    \caption{Mass exclusions (in \textit{orange}) corresponding to 95\% confidence level for the $I_{\tilde{q}_{L/R}\rightarrow\tilde{B}}$ cascade with $\lambda^{''}_{113}$ (\textit{left}) and $I_{\tilde{b}_{1}\rightarrow\tilde{B}}$ cascade with $\lambda^{''}_{112}$ (\textit{right}). }
    \label{fig:bino_lsp_st_113}
\end{figure}

As mentioned in Sec. \ref{ssec:cascade}, in some cases the scan 
results obtained from \texttt{CheckMATE\;2} do not yield a 
well-defined exclusion contour. In 
Fig.\,\ref{fig:bino_lsp_st_113} we show the scan results as 
scatter plots for the $I_{\tilde{q}_{L/R} \to \tilde{B}}$ 
cascade with $\lambda^{''}_{113}$ and $I_{\tilde{b}_{1}\to 
\tilde{B}}$ cascade with $\lambda^{''}_{112}$. 

\begin{figure}[hbt!]
    \centering
     \includegraphics[width=0.86\linewidth]{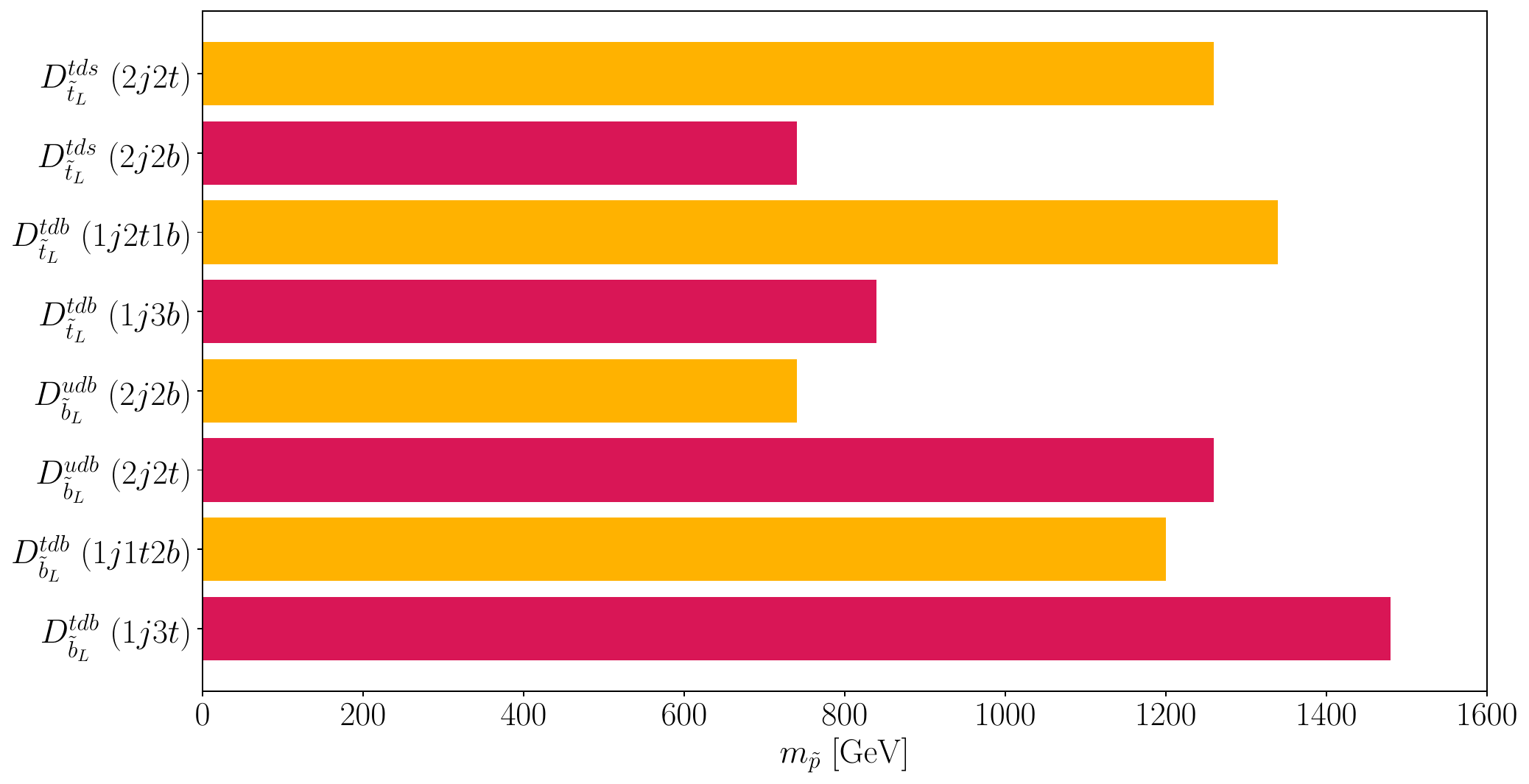}
    \caption{Direct production limits for the third generation squarks ($\tilde{t}_L$, $\tilde{t}_R$, $\tilde{b}_L$ and $\tilde{b}_R$) assuming 100\% branching to each decay mode. The parentheses after each benchmark $D_{\tilde{p}}$ show the specific branching mode.}
    \label{fig:direct_tRbR_branched}
\end{figure}

In Fig.\,\ref{fig:direct-lsp}, we 
show the results for the benchmark scenarios provided in Tables\,\ref{tab:gluino_UDD}-\ref{tab:nv2slep_UDD}. However, in 
some cases, as shown in Table\,\ref{tab:squark_UDD}, the third generation squarks can have multiple decay modes for the same UDD coupling. The results shown in Fig.\,\ref{fig:direct-lsp} 
consider 50\% branching to each decay mode. The exact branching 
fraction depends on the details of the mass spectrum of the intermediate sparticles. For completeness, Fig.\,\ref{fig:direct_tRbR_branched} shows results assuming 100\%
branching fraction for each decay mode, wherever multiple decay modes are possible. These results are obtained using \texttt{CM2}.

\providecommand{\href}[2]{#2}\begingroup\raggedright\endgroup

\end{document}